\documentclass[prl, superscriptaddress, twocolumn]{revtex4-2}
\usepackage{braket}
\usepackage{xcolor}
\usepackage{graphicx}
\usepackage{dcolumn}
\usepackage{bm}
\usepackage{svg}
\usepackage{float}
\usepackage{amsmath}
\usepackage{upgreek}
\usepackage{natbib}
\usepackage{hyperref}
\usepackage{graphicx}
\hypersetup{
    colorlinks=true,
    linkcolor=blue,
    citecolor=blue,
    filecolor=blue,      
    urlcolor=blue,
    pdftitle={Overleaf Example},
    pdfpagemode=FullScreen,
    }
\graphicspath{ {Images/} }

\begin{document}

\preprint{APS/123-QED}

\title{Near-video frame rate quantum sensing using Hong-Ou-Mandel interferometry}

\author{Sandeep Singh}
\email{sandeep@prl.res.in}
\affiliation{Photonic Sciences Lab., Physical Research Laboratory, Navrangpura, Ahmedabad 380009, Gujarat, India}
\affiliation{Indian Institute of Technology-Gandhinagar, Ahmedabad 382424, Gujarat, India}

\author{Vimlesh Kumar}
\affiliation{Photonic Sciences Lab., Physical Research Laboratory, Navrangpura, Ahmedabad 380009, Gujarat, India}
 
\author{Varun Sharma}
\affiliation{Photonic Sciences Lab., Physical Research Laboratory, Navrangpura, Ahmedabad 380009, Gujarat, India}
 
\author{Daniele Faccio}
\affiliation{School of Physics and
Astronomy, University of Glasgow, Glasgow G12 8QQ,
UK}
 
\author{G. K. Samanta}
\affiliation{Photonic Sciences Lab., Physical Research Laboratory, Navrangpura, Ahmedabad 380009, Gujarat, India}%

\date{\today}

\begin{abstract}
\noindent 
Hong-Ou-Mandel (HOM) interference, the bunching of two indistinguishable photons on a balanced beam-splitter, has emerged as a promising tool for quantum sensing. There is a need for wide spectral-bandwidth photon pairs (for high-resolution sensing) with high brightness (for fast sensing). Here we show the generation of photon-pairs with flexible spectral-bandwidth even using single-frequency, continuous-wave diode laser enabling high-precision, real-time sensing. Using 1-mm-long periodically-poled KTP crystal, we produced degenerate, photon-pairs with spectral-bandwidth of 163.42$\pm$1.68 nm resulting in a HOM-dip width of 4.01$\pm$0.04 $\upmu$m to measure a displacement of 60 nm, and sufficiently high brightness to enable the measurement of vibrations with amplitude of $205\pm0.75$ nm and frequency of 8 Hz. Fisher-information and maximum likelihood estimation enables optical delay measurements as small as 4.97 nm with precision (Cramér-Rao bound) and accuracy of 0.89 and 0.54 nm, respectively, therefore showing HOM sensing capability for real-time, precision-augmented, in-field quantum sensing applications.
\end{abstract}


\maketitle

\section{Introduction}
\noindent
Quantum sensing and metrology is a rapidly growing field due to its outstanding features in measuring physical parameters with great precision and accuracy, outperforming the pre-existing technologies based on the principles of classical physics \cite{2018QS}. In recent years, quantum sensing has enabled the measurement of several key physical quantities, including the measurement of the electrical field \cite{bian2021electricnanoscale}, magnetic field \cite{tsukamoto2022magneticaccurate}, vacuum \cite{NISTquantumsensor}, temperature \cite{Tempsensor}, and pressure \cite{pressuresensor} with unprecedented precision and accuracy. Among various quantum sensors as presented in Ref. \cite{RevModPhys.89.035002}, Hong-Ou-Mandel (HOM) interferometry, since its discovery \cite{HOM.PhysRevLett.59.2044}, has found a large number of applications within quantum optics for a variety of advantages, including easy development and implementation, sensitivity only to photon group delay and not to phase shifts \cite{dispersioncancellation}. \\
\indent
HOM interference, purely a non-classical phenomenon, is observed for two photons that are identical in all degrees of freedom, including spin, frequency, and spatial mode, when simultaneously entering on a lossless, 50:50 beam splitter through different input ports bunch together into one of the output ports. The coincidence probability between the two output ports shows a characteristic low or zero coincidence, known as the HOM interference dip, directly related to the level of indistinguishability or the degree of purity of the photons \cite{2010accessing}. As a result, the change in coincidence counts can be treated as a pointer to estimate the time delay between photon properties, forming the basis of the HOM interferometer-based quantum sensors to sense any physical process influencing the photon delay. As such, efforts have been made to use the HOM-based quantum sensors to characterize the single photon sources \cite{Cassemiro2010}, precise measurement of time delays between two paths \cite {HOM.PhysRevLett.59.2044, lyons2018attosecond, chen2019biphotonbeat}, characterization of ultrafast processes \cite{lipka2021}, measurement of frequency shifts \cite{frequencyshift, Photongyroscope} and the spatial shift \cite{quantummicroscopy}. 
On the other hand, the use of Fisher Information (FI), a measure of the mutual information between the interfering photons, has enabled the maximum precision to access a lower limit decided by the Cramér-Rao bound \cite{Rao1945,cramer1946}. In fact, the use of peak FI of the HOM interferometer, the point of maximum sensitivity, has enabled five orders of magnitude enhancement in the resolution of time delay measurements \cite{lyons2018attosecond,chen2019biphotonbeat,harnchaiwat2020tracking}. A typical FI-based measurement approach involves tuning the HOM interferometer for the peak FI point and performing iterative measurement to estimate any optical delay between the two photons. However, the number of iterations/measurements and hence the total time needed to reach the targeted precision decreases with the increase of FI value. On the other hand, the FI value can be increased by optimizing the parameters affecting the visibility and the width of the HOM interference dip. Despite the experimental demonstration of nanometer path length (few-attosecond timing) precision \cite{lyons2018attosecond,kwait2023} and a recent theoretical study on the tailoring of the spectral properties \cite{timinglimits, tamma2023} of photon pairs to achieve precision beyond the value reported in the laboratory condition, the width of the HOM interference dip governed by the bandwidth of the pair photons remained the limiting factor for optimal measurement precision \cite{lyons2018attosecond, chen2019biphotonbeat, Beyondcoincidence, Noiselimit} in HOM interferometer based dynamic (real-time) or fast sensing applications. As such, ultra-broadband photon sources have long been hailed as a vital prerequisite for ultra-precise HOM interferometry. On the other hand, the dynamic sensing process requires the single-photon source to have high brightness. On the other hand, a new strategy has been reported very recently where entangled photons have been used to produce the HOM beating effect at very different frequencies (visible and infrared) \cite{kwait2023}. As the frequency of the beating is proportional to the frequency detuning of the entangled photons \cite{kwait2023, beatnote_biphoton} the HOM beating, in conjunction with FI analysis, enabled the measurement of low-frequency (0.5 - 1 Hz) vibration with a precision of $\sim$2.3 nm  using low (18000) number of photons pairs \cite{kwait2023}\\
\indent
Nonlinear spontaneous parametric down-conversion (SPDC) \cite{klyshko1967coherent, boyd2020nonlinear} process-based sources, where the annihilation of a high energy pump photon produces two daughter photons simultaneously, have evolved as a workhorse for quantum optics experiments, including HOM based sensors. However, the development of high brightness single-photon source requires a nonlinear crystal with high figure-of-merit (FOM) \cite{ebrahimzadeh2001optical}, long crystal length, and high intensity of the pump beam. As such, periodically poled nonlinear crystals can be used due to their high intrinsic FOM. However, the increase in the crystal length reduces the spectral bandwidth of the single-photon source. On the other hand, the increase of input pump intensity and spectral bandwidth of the generated photons using ultrafast laser \cite{okano201500.5micron} has its own limitations in terms of expensive and bulky system architecture for any practical on-field uses. Therefore, it is imperative to explore a continuous-wave (CW) pump-based bright single-photon source with high brightness and spectral bandwidth for ultra-precise HOM interferometry. \\
\indent
Recently, we reported a CW pumped high brightness, degenerate single-photon source at 810 nm using periodically-poled potassium titanyl phosphate (PPKTP) crystal in non-collinear, type-0, phase-matched geometry \cite{Jabir:17, singh2023}. Using the same phase-matching geometry, here, we report on the experimental demonstration showing the dependence of spectral bandwidth of degenerate SPDC photons and subsequent change in the HOM interference dip width on the length of the PPKTP crystal. Pumping the PPKTP crystal of length 1 mm, using a single-frequency, CW diode laser at 405.4 nm we observed the spectral width of the SPDC photons to be as high as 163.42 $\pm$ 1.68 nm producing a HOM interference dip having a full-width-at-half-maximum (FWHM) as small as 4.01 $\pm$ 0.04 $\upmu$m. Using such small HOM interference dip width, we have dynamically measured the threshold vibration amplitude corresponding to the optical delay (twice the vibration amplitude of the piezo mirror) introduced between the photons as small as 205$\pm$0.75 nm at a frequency as high as 8 Hz. On the other hand, the reduction of crystal length from 30 mm to 1 mm shows a 17 times enhancement in the magnitude of the peak FI value and achieves a targeted precision (say $\sim$5 nm or $\sim$16.7 attosecond optical delay) for a number of iterations/measurements (or the total experiment time) as low as 3300 (19 minutes).\\
\section*{Theoretical Background}\label{Background}
\noindent
The spectral brightness of SPDC photons generated from a nonlinear crystal of length, $L$ can be expressed as follows,
\begin{equation}\label{eq:1}
I_{SPDC} \sim \mbox{sinc}^{2}\left( \Delta kL/2 \right) 
\end{equation}
where $\Delta k$ is the momentum mismatch among the interacting photons (pump and down-converted pair-photons) given as follows
\begin{equation}\label{eq:2}
\Delta k = k_{p} - k_{s} -k_{i} -\frac{2\pi m}{\Lambda}
\end{equation}
Here, $k_{p,s,i}$ = $\frac{2 \pi n_{p,s,i}}{\lambda_{p,s,i}}$ is the wavevector and $n_{p,s,i}$ is the refractive index of the nonlinear crystal for the pump, $\lambda_{p}$, signal, $\lambda_{s}$, and idler, $\lambda_{i}$, wavelengths. $\Lambda$ is the grating period of the periodically-poled nonlinear crystal to satisfy the quasi-phase-matching (QPM) condition. The odd integer number, $m$, is the QPM order. For maximum efficiency, we consider $m$ = 1. The momentum mismatch between degenerate photons around the central angular frequency, $\omega_{p}/2$, can be represented as \cite{RDas:09, chen2021broadband}
%
\begin{equation}\label{eq:3}
\Delta k = \Delta k |_{\omega=\omega_{p}/2}+M|_{\omega=\omega_{p}/2}{\Delta\omega} + \frac{1}{2}K |_{\omega=\omega_{p}/2} {\Delta\omega}^2 + ....
\end{equation}
%
Assuming the QPM is achieved at degenerate SPDC photons with angular frequencies of pump, signal, and idler as $\omega_{p}$, $\omega_{s}$, and $\omega_{i}$, respectively, for the grating period, $\Lambda$, the first term of the right-hand of Eq. (\ref{eq:3}) can be made zero. In this situation, owing to the energy conservation, $\omega_{p}$ = $\omega_{s}$ + $\omega_{i}$, the spectral bandwidth of the down-converted photons for the fixed pump frequency defined as $\Delta\omega$ = $\Delta\omega_{s}$ = -$\Delta\omega_{i}$ =$\omega-\omega_{p}/2$, is decided by the second term, M = $[\partial k_{s}/\partial \omega - \partial k_{i}/\partial\omega]_{\omega = \omega_{p/2}}$, known as group-velocity mismatch. However, for degenerate $(\omega_{s} = \omega_{i} = \omega_{p}/2)$, type-0 SPDC process, the group-velocity mismatch term vanishes or has negligible value. Under this condition, the spectral acceptance bandwidth of the SPDC photons is essentially determined by the much smaller third term of Eq. (\ref{eq:3}), commonly known as group-velocity dispersion with a mathematical form $K = \partial^2 k/\partial \omega^2|_{\omega=\omega_{p/2}}$. 
Under this condition, the spectral bandwidth of the SPDC photons can be written as, 
\begin{equation}\label{equ:4}
    \Delta\omega \propto \frac{1}{\sqrt{KL}}
\end{equation}
Thus, one can utilize the length of the nonlinear crystal as the control parameter in the SPDC process to generate paired photons of broad spectral bandwidth near the degeneracy.
On the other hand, it is well known that the optical delay range or the width, $\sigma$, of the HOM interference is proportional to the coherence length of the photon wave packets \cite{dispersioncancellation,abouraddy2002QOCT} or the ensemble dephasing time of SPDC photons {\cite{BSham2022coherence} bunching on the beam splitter. Although one can find the width, $\sigma$, of the HOM interference dip from the Fourier transform of the spectral density function of the paired photons \cite{2020spectrallyresolved, PhysRevLett.91.083601}, we used the coherence length, $L_{c}$, of the photon wave packets and Eq. (\ref{equ:4}) to find the dependence of HOM interference dip width on the crystal length as, 
\begin{equation}\label{equ:5}
\sigma \hspace{5pt} \sim \hspace{5pt} L_{c} \hspace{5pt} \propto \hspace{5pt} 2\pi c\sqrt{KL}
\end{equation}
where c is the speed of light in vacuum. It is evident from Eq. (\ref{equ:5}) that the width, $\sigma$, of the HOM interference dip can be controlled by simply varying the crystal length even in presence of the single-frequency, CW diode laser as the pump. 

\section*{Experiment}\label{Experiment}
\noindent
The schematic of the experimental configuration is shown in Fig. \ref{Exp}. The details of the experiment can be found in the Methods section. A single-frequency, CW diode laser providing 20 mW of output power at 405.4 nm is used to create downconverted photons in the periodically poled KTiOPO$_{4}$ (PPKTP) crystals of 1 $\times$ 2 mm$^2$ aperture but of five different interaction lengths, $L$ = 1, 2, 10, 20, and 30 mm. All the crystals have a single grating period of $\Lambda$ = 3.425 $\upmu$m corresponding to the degenerate, type-0 ($e $→$ e + e$) phase-matched parametric down-conversion (PDC) of 405 nm at 810 nm. 
%
\begin{figure}[h]
\centering
\includegraphics[width=\linewidth]{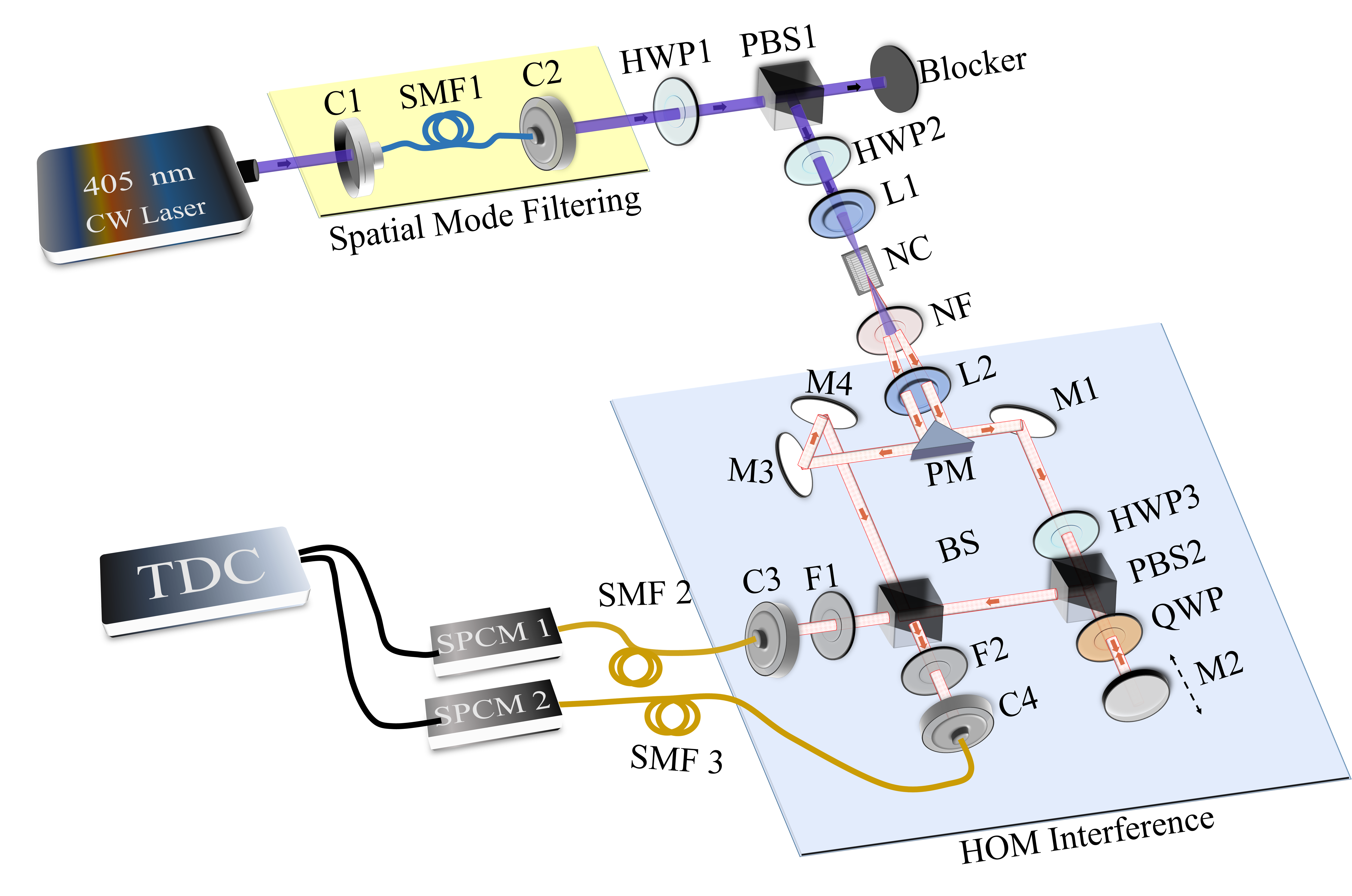}
\caption{Schematic of the experimental setup. \textbf{C1-4}: fiber coupler, \textbf{SMF1-3}: single-mode fiber, \textbf{HWP1-2}: half-wave plate, \textbf{PBS1-2}: polarizing beam splitter cube, \textbf{L1-2}: lenses, \textbf{NC}: PPKTP crystal of grating period = 3.425 $\upmu$m, \textbf{NF}: notch filter, \textbf{PM}: prism mirror, \textbf{M1-4}: dielectric mirrors, \textbf{QWP}: quarter wave plate, \textbf{BS}: 50:50 beam splitter cube, \textbf{F1,2}: high-pass filter, \textbf{SPCM1-2}: single photon counting module, \textbf{TDC}: time-to-digital converter.}
\label{Exp}
 \end{figure}
\noindent
The noncollinear down-converted photon pairs having annular ring spatial distribution with signal and idler photons on diametrically opposite points are guided in two different paths and finally made to interact on the two separate input ports of a balanced beam splitter (BS). The combination of the piezo-electric actuator (PEA) (NF15AP25), placed on a motorized linear translation stage (MTS25-Z8), is used to provide both fine and coarse optical delay. The photons from the output ports of the BS, after being extracted by long pass filters (F1, F2), are coupled into the single mode fibers, SMF2 and SMF3, through the fiber couplers, C3, and C4, respectively. The collected photons are detected using the single-photon counting modules, SPCM1-2 (AQRH-14-FC, Excelitas), and counted using the time-to-digital converter (TDC). All the data is recorded at a typical pump power of 0.25 mW with a coincidence window of 1.6 nanoseconds.\\

\section*{Results and Discussions}

\subsection*{Displacement sensing using HOM interferometer} 
\noindent
First, we have characterized the HOM interference for different crystal lengths and calibrated various system parameters as presented in the Materials and Methods section. Knowing the dependence of the sensitivity on the crystal length, we have studied the performance of the HOM interferometer to measure the static displacement of the mirror, introducing an optical delay between the photons. Using the 1 mm long PPKTP crystal resulting in HOM interference dip width of 4.01 $\pm$ 0.04 $\upmu$m, we have moved the mirror using the motorized stage to a displacement of $\sim$ 1.25 $\upmu$m away from the zero-optical delay position. This translation resulted in a positive optical delay of $\sim$ 2.5 $\upmu$m corresponding to the region having maximum sensitivity of the HOM region of the sensor. We moved the mirror in a step of 60 nm (resulting optical delay of 200 attoseconds) and measured the coincidence counts. The results are shown in Fig. \ref{static stability}. It is evident from Fig. \ref{static stability}(a), the displacement of the mirror by 60 nm results in the change of the coincidence counts of more than 50 per 20 ms of integration time. Such a large change in the coincidence count, much higher than the dark count and accidental counts, confirms the possibility of measurement of static displacement as low as 60 nm.\\
\begin{figure}[h]
\centering
\includegraphics[width=\linewidth]{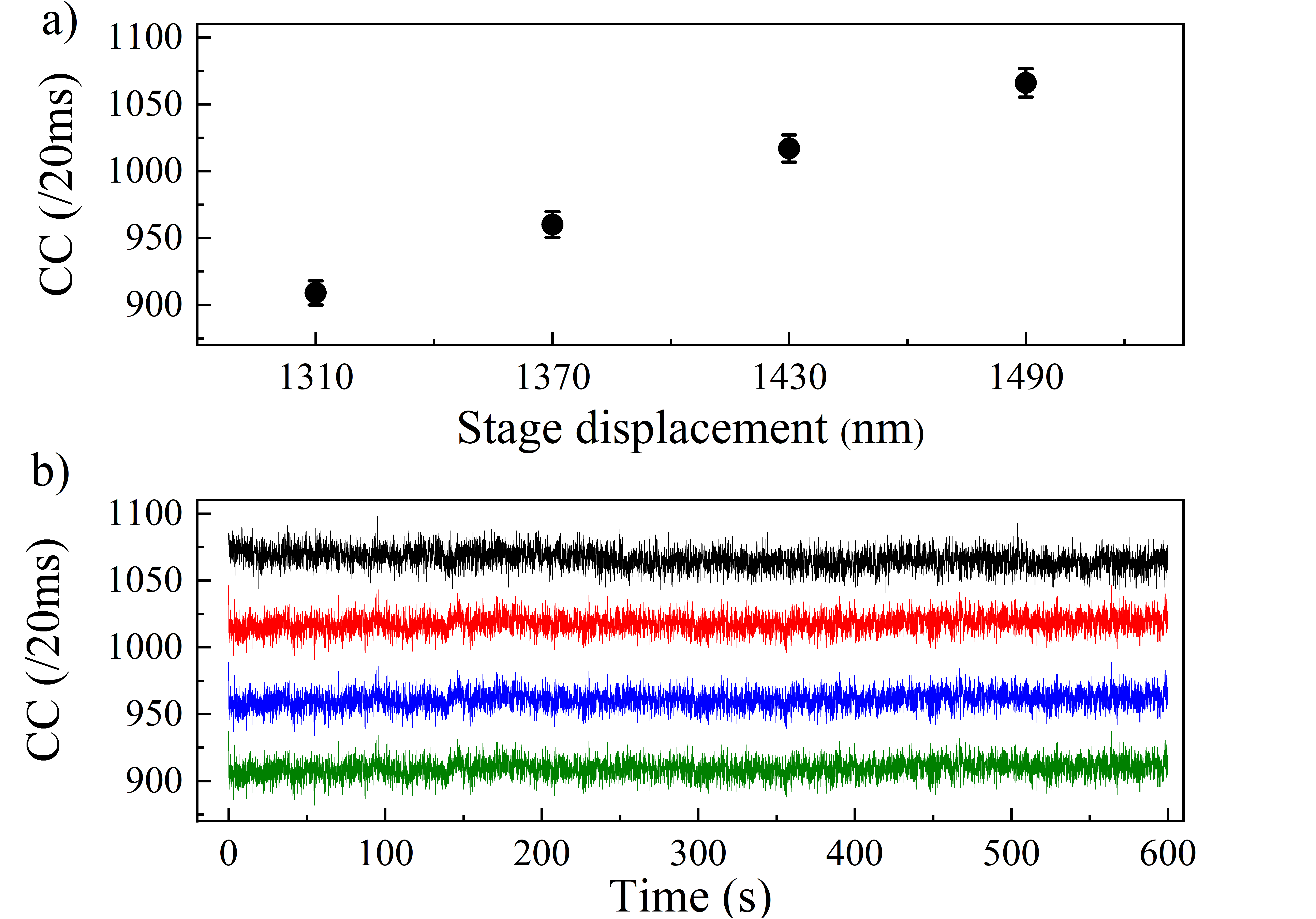}
\caption{a. Variation in the coincidence counts (CC) with nanometer displacement of the optical delay mirror, M2. b. Temporal stability of coincidence counts at different positions of the delay mirror, M2.}
\label{static stability}
\end{figure}
%
Such a change in the coincidence count is easily detectable, confirming the displacement measurement as small as 60 nm. Further, to confirm the reliability of the measurement, we measured the temporal variation of the coincidence counts over 10 minutes for each mirror position. It is evident from Fig. \ref{static stability}(b), that the temporal variation (with standard deviation $\sim$ 7.4$\%$) of the coincidence counts due to various parameters, including the laser intensity fluctuation, air current, and local temperature instability in the laboratory, is much smaller than the change in the coincidence counts due to the static displacement of the mirror. The possibility of smaller static displacement can be possible by suitably reducing the fluctuation of the coincidence counts in the experiment.\\  

\subsection*{Dynamic vibration sensing using HOM interferometer} 
\noindent 
Having the calibration data (see Materials and Methods) for the piezo for static displacement, we have studied the performance of the HOM interferometer for a dynamic signal. In doing so, we have used the HOM interference dip width (FWHM) of 4.01 $\upmu$m and adjusted the initial positive delay by a value of $\sim$3 $\upmu$m. Driving the PEA with a periodic voltage signal in a triangular wavefront of controllable frequency and amplitude, we have recorded the temporal variation of the coincidence counts of the HOM with the results shown in Fig. \ref{dynamic sensing}. As evident from Fig. \ref{dynamic sensing}(a), the vibrations of the PEA with a periodic triangular signal of period 1 Hz and the peak-to-peak voltage of 1 V corresponding to a peak-to-peak optical delay of $\sim$1.0 $\upmu$m, as estimated from the calibration data (see Materials and Methods), produces a peak-to-peak variation in the coincidence counts of $\sim$400 over the mean value of 960 at a period of 1 Hz, the same as the driving frequency. Further, we changed the peak-to-peak vibration amplitude of PEA from the calibration data and measured the peak-to-peak optical delay from the coincidence counts with the results shown in Fig. \ref{dynamic sensing}(b). We have repeated this exercise twice and termed the results as set-1 (red data points) and set-2 (blue data points). As evident from Fig. \ref{dynamic sensing}(b), the peak-to-peak optical delay (which is twice the displacement of the PEA) measured using the change in the coincidence counts of HOM exactly follows the set delay at a slope of 0.835 $\pm$ 0.004. Ideally, the slope should have a value of 1 to maintain one-to-one correspondence of the set and measured values. However, the discrepancy between the experimental slope with respect to the ideal value can be attributed to the error in the coupling of the applied voltage to the PEA and its hysteresis \cite{zhou2012piezo,grzybek2022creep}. While increasing the PEA vibration amplitude through the applied voltage signal, we realized that the threshold (to replicate the shape of the input waveform exactly) peak-to-peak vibration amplitude, which can be measured dynamically using this technique, is found to be $205\pm0.75$ nm. Subsequently, the average maximum attainable resolution between two consecutive vibration amplitudes is found to be $\sim$80 nm. However, one can access smaller peak-to-peak vibration amplitude ($<$50 nm) (without replicating the shape of the input wavefront) from the peak-to-peak variation of the coincidence counts. \\
\begin{figure}[h]
\centering
\includegraphics[width=\linewidth]{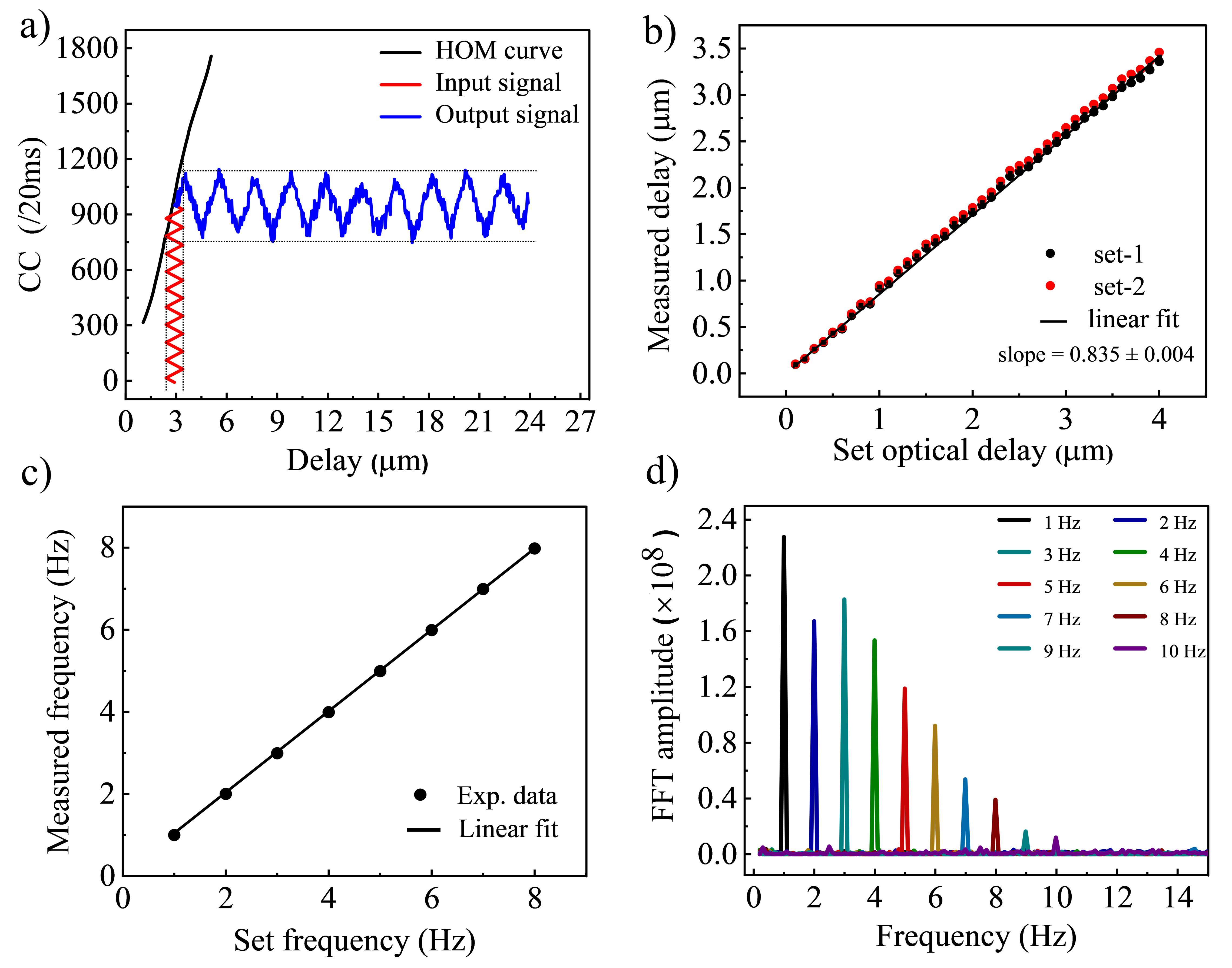}
\caption{Vibration measurement using HOM interferometer sensor. a. Variation of coincidence counts for periodic triangular voltage signal of peak-to-peak amplitude 1 V at 1 Hz applied to the PZT stage. b. Variation of PEA displacement, and c. driving frequency measured from the coincidence counts with respect to the set values. d. Fast Fourier transformation of the time-varying coincidence counts measuring the driving frequency of the applied voltage signal.}
\label{dynamic sensing}
\end{figure}

\indent
As the HOM interferometer deals with single photons, it is essential to integrate the detected coincidence events for a substantial time to get a tangible number of coincidence counts while making any measurement. Again the required integration time is highly influenced by the brightness of the single-photon source. In the current experiment, we used high brightness single photon source based on PPKTP crystal \cite{Jabir:17,singh2023}. As a result, we can keep the integration time as low as a few tens of milliseconds. Here, we have also tested the HOM interferometer to measure the vibration frequency of the PEA. Keeping the experimental parameters the same as previous, we have changed the frequency of the triangular voltage signal to the PEA at the constant peak-to-peak voltage of 1 V corresponding to the optical delay of $\sim$1 $\upmu$m. While we adjusted the set frequency from the function generator, we recorded the temporal variation of the coincidence counts for each driving frequency and performed the fast Fourier transform (FFT) to experimentally measure the driving frequency. As evident from Fig. \ref{dynamic sensing}(c), the measured frequency using the coincidence data exactly matches the driving frequency. The linear fit (line) to the experimental data (dots) shows a slope of 1, confirming the reliable measurement of the frequency of the driving field. However, we have restricted our measurement to 8 Hz as the FFT signal amplitude, despite having the peak at the driving frequency, as shown in Fig. \ref{dynamic sensing}(d), decreases with the increase of the driving frequency. Such a decrease can be understood as follows. In FFT analysis, the accuracy depends on the number of acquired samples from the experimental data. For example, the higher resolution in estimating the driving frequency using FFT analysis requires many data points. Again, the number of samples/data points can be increased by increasing the sampling rate. However, in the present case, the sampling rate is restricted to 50 Hz due to the requirement of a high exposure/integration time of 20 milliseconds limited by the generation rate of the SPDC source and the processing speed of the TDC. Although we have 50 data points to perform FFT analysis for a driving frequency of 1 Hz, the available data points vary inverse to the driving frequency as $M = 50/f$, where $M$ is the number of samples available for FFT analysis and $f$ is the driving frequency. As a result, it becomes difficult to retrace the waveform of the input signal at an increased frequency which leads to the gradual decrease in the FFT signal amplitude (decrease in signal-to-noise, SNR, ratio) and the subsequent appearance of its second harmonic peak. Further, an increase in the driving frequency requires a decrease in data integration time (to achieve a higher sampling rate) through the increase of the photon generation rate of the source and the use of TDC with low data acquisition latency. However, in the best-case scenario, one can, in practice, use HOM interferometer-based sensor to measure dynamic vibration signals of unknown frequency restricted to a few tens of hertz only.\\
\indent
We further study the performance of the HOM interferometer-based quantum sensor to measure any arbitrary vibration signal. Considering the initial position on the high-sensitivity region of the HOM interference dip as the zero displacement point, we drove the PEA with an external voltage signal to the PEA in the triangular waveform of varying amplitude (peak-to-peak voltage range of 0.7 to 2 V) and frequency (1 Hz to 8 Hz). The results are shown in Fig. \ref{random signal response}. As evident from Fig. \ref{random signal response}(a), the displacement of the PEA measured from the calibration curve (see Materials and Methods) shows a temporal variation in both amplitude and frequency, confirming the arbitrary vibration. However, as evident from Fig. \ref{random signal response}(b), the variation of the coincidence counts recorded for an integration time of 20 ms exactly follows the applied signal. Such observation confirms the possibility of measurement of any arbitrary vibration, for example, low amplitude and frequency seismic S- and P-waves, using a HOM interferometer-based quantum sensor.\\
\begin{figure}[h]
\centering
\includegraphics[width=\linewidth]{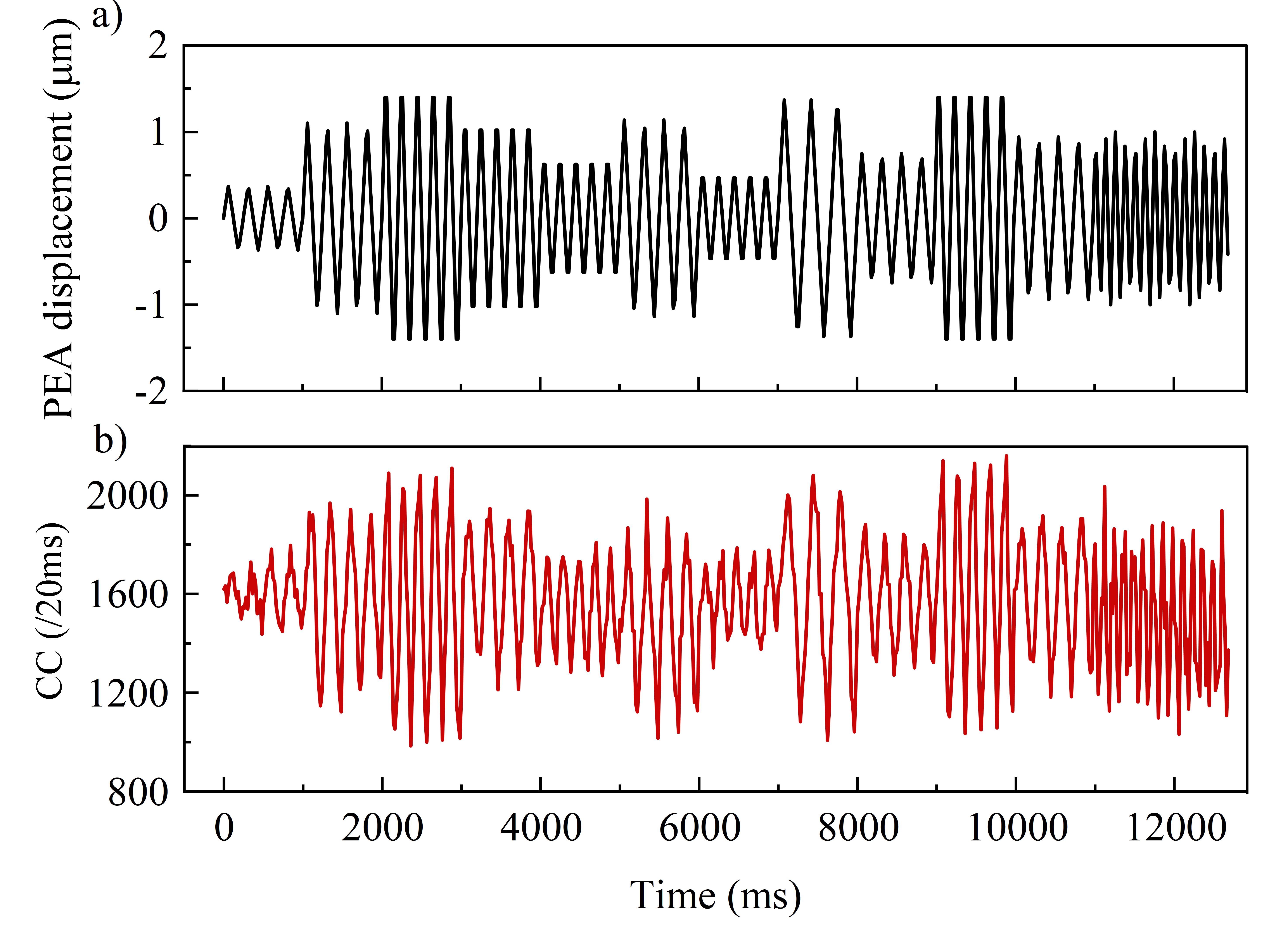}
\caption{a. Temporal variation of PEA displacement and b. corresponding variation in the coincidence counts of the HOM interferometer for the external vibration.}
\label{random signal response}
\end{figure}
\indent 
The current study establishes the potential of the HOM interferometer as a quantum sensor for real-time detection of time-varying signals in the frequency ranging from 1 Hz to 8 Hz and any physical process producing optical delays as low as $0.41 \pm 0.06$  $\upmu$m with a precision of $\sim$0.19 $\upmu$m. While the frequency range can further be increased to tens of hertz by reducing the exposure time of the measurement, the precision in the real-time measurement of lower optical delay is fundamentally limited by the statistical and systematic error, especially at lower variation in coincidence counts \cite{fox2006quantum}. As such, one needs to explore the statistical concept of Fisher information to attain a precision that has a lower limit governed by the Cramér-Rao bound \cite{lyons2018attosecond, harnchaiwat2020tracking, chen2019biphotonbeat}.\\
\section*{Precision augmented sensing}\label{Precision}
\noindent
The ultimate limit on the precision in measurement is decided by the Cramér-Rao bound \cite{Rao1945,cramer1946} defined as,
\begin{equation}\label{eq:6}
Var(\tilde{x}) \hspace{5pt} \geq \hspace{5pt} \frac{1}{NF(x)}
\end{equation}
where $\tilde{x}$ represents an unbiased estimator to estimate the physical parameter and var($\tilde{x}$) is the variance in the measurement. $F(x)$ represents the Fisher information (FI), a metric describing the amount of information available from a probability distribution of an unknown parameter, for estimating the information of the parameter $x$ in a single measurement, and N is the total number of measurements performed in the experiment. In the case of HOM interference, the mathematical form of FI is primarily determined by the temporal mode profile of the signal and idler photons \cite{lyons2018attosecond}. As the down-converted photons generated through the type-0 SPDC process have a Gaussian spectral profile, the FI can be modeled as,
\begin{equation}\label{eq:7}
F =\frac{4s^2\alpha^2(\gamma-1)^2(1+\gamma)}{(e^{s^2}-\alpha)(\alpha-\alpha\gamma+e^{s^2}(1+3\gamma))\sigma^2}
\end{equation}\vspace{1pt}\\
here, $\alpha$, $\gamma$, and $\sigma$ represent the HOM interference visibility, the rate of loss in photon detection, and the FWHM width of the HOM curve, respectively. $s=\frac{x}{\sigma}$ is the parameter controlling the distinguishability of the photon arriving at the beam splitter.
\indent
As evident from Eq. (\ref{eq:7}), keeping all parameters constant, one can increase the magnitude of FI, $F$ by employing the HOM interference of smaller dip width ($\sigma$). Again, it is evident from Eq. \ref{equ:5} that for a given laser parameter, the dip width of the HOM interference varies with the square root of the crystal length. Since the crystal length, a physical parameter controllable in the experiment, using Eq. (\ref{equ:5}) in Eq. (\ref{eq:7}), we can derive a general expression for FI in terms of crystal length, L, as,
\begin{equation}\label{eq:8}
F =\frac{4x^2\alpha^2(\gamma-1)^2(1+\gamma)}{A^4(e^{\frac{x^2}{A^2L}}-\alpha)(\alpha-\alpha\gamma+e^{\frac{x^2}{A^2L}}(1+3\gamma))L^2}
\end{equation}\vspace{1pt}\\
Here, A is a constant determined by the crystal properties. It is evident from Eq. (\ref{eq:6}) that for a fixed number of measurements/iterations, $N$, the precision can be enhanced, or for a fixed precision, the number of measurements/iterations, $N$, can be reduced with the increase of the magnitude of $F$ while saturating the Cramér-Rao bound. Therefore, Eq. (\ref{eq:8}) sets the foundation for our further study of precision augmented quantum sensing.\\ 
\begin{figure}[h]
\centering
\includegraphics[width=\linewidth]{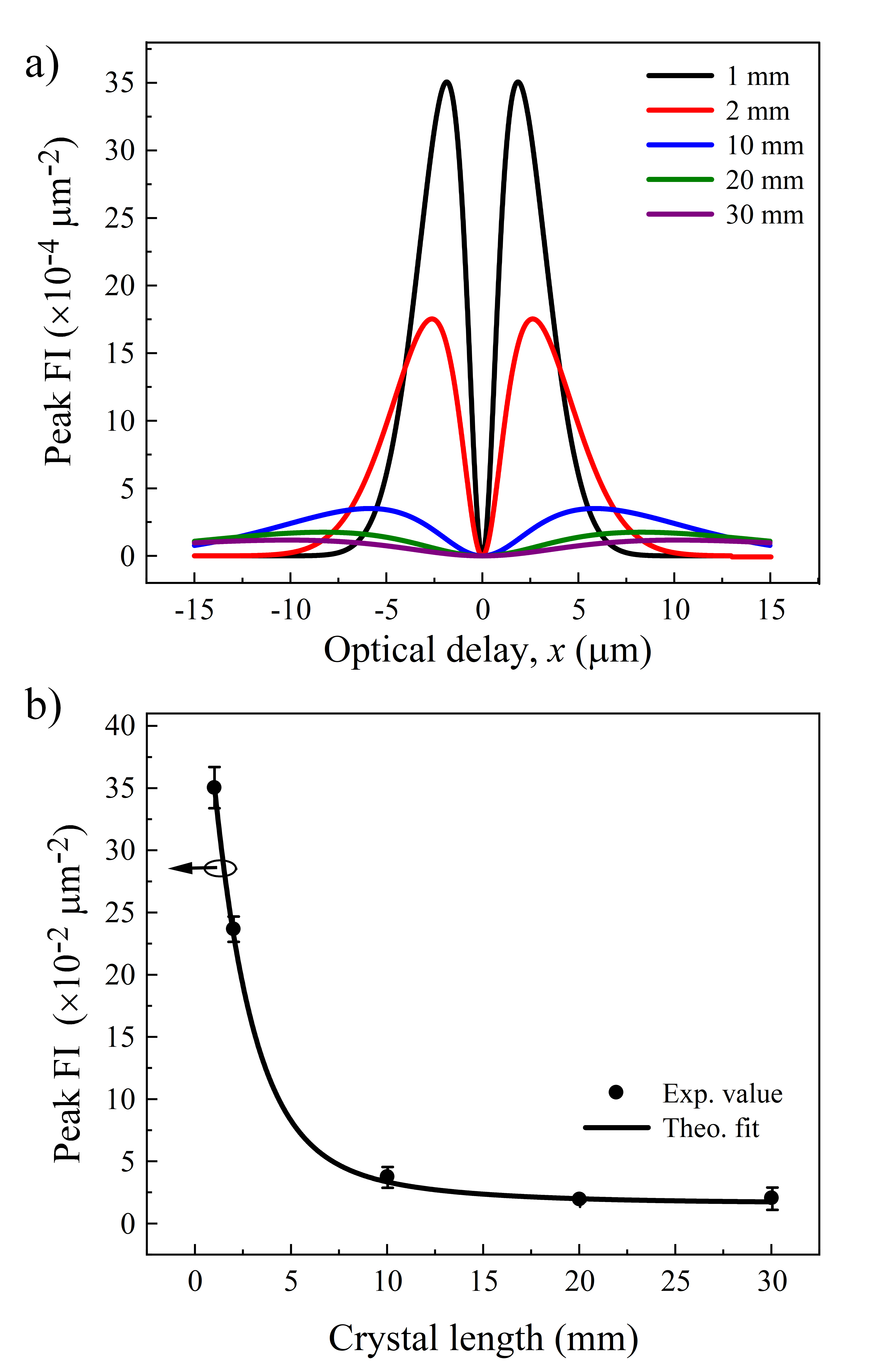}
\caption{a. Variation of Fisher information as a function of optical delay, $x$, between the signal and idler photons for different crystal lengths. b.  Dependence of the peak FI value with the length of the nonlinear crystal. Solid line is the theoretical fit (Eq. \ref{eq:8}) to the experimental data.}
\label{FI vs crystal length}
\end{figure}
\indent
To get a better perspective on the dependence of FI on the crystal length, we have plotted Eq. (\ref{eq:8}) as the function of optical delay, $x$ while keeping the experimentally measured values of $\alpha$ and $\gamma$ as constant. We have used crystal lengths, $L$ = 1, 2, 10, 20, and 30 mm, as available in our lab. The results are shown in Fig. \ref{FI vs crystal length}(a). As evident from Fig. \ref{FI vs crystal length}(a), the FI has the usual double-peak Gaussian profiles for HOM visibility $\alpha < 1$. It has been previously observed \cite{lyons2018attosecond} that the HOM visibility, $\alpha$  influences the magnitude and the separation between two peaks of FI, respectively. However, in the current experiment, we observed that for a fixed value of HOM interferometer visibility, the shape and peak value of FI depend on the value of HOM interference dip width, $\sigma$. As evident from Fig. \ref{FI vs crystal length}(a), the decrease in the width of the HOM interference dip due to the decrease of crystal length (see Eq. (\ref{equ:5})) not only reduces the separation of the two peaks of FI but also show a substantial increase in their respective peak values. To put things in perspective, using the experimental parameters in Eq. (\ref{eq:8}), we estimated the peak values of FI for crystal lengths 1, 2, 10, 20, and 30 mm. The results are shown in Fig. \ref{FI vs crystal length}(b).\\
\noindent
It is evident from Fig. \ref{FI vs crystal length}(b) that the peak value of FI (black dots) decreases from {35.06 $\pm$ 1.65) x $10^{-4}$ $\upmu$m$^{-2}$ to (2.07 $\pm$ 0.19) x $10^{-4}$ $\upmu$m$^{-2}$ for the increase of the crystal length from 1 mm to 30 mm due to the corresponding increase in the FWHM width of the HOM interference dip from 4 $\upmu$m to 21.2 $\upmu$m (see Fig. \ref{crystal len vs spectral bandwidth}). Such observation clearly shows the enhancement in peak value of FI (here 17$\times$) by simply reducing the nonlinear crystal length (here 1/30) even using a single-frequency diode laser. It is also interesting to note that the current peak value of FI shows a $\sim$24$\times$ enhancement as compared to the FI value obtained using an ultrafast laser \cite{lyons2018attosecond}, with the further possibility of enhancement by using  an ultrafast laser generating SPDC photons in a thin nonlinear crystal. Such an increase in the peak value of FI can be useful in faster saturation of the Cramér-Rao bound as defined in Eq. (\ref{eq:6}) at the lower number of measurements/iterations to achieving a predefined precision in the measurement of optical delay. 
\subsection*{FI based measurement}
\noindent
The precise estimation of a physical parameter such as optical delay, $x$, can be obtained through the maximum likelihood estimator \cite{lyons2018attosecond}. In the current study, the estimator of the maximum likelihood function  for measuring the optical delay can be defined as,
\begin{equation}\label{eq:9}
\tilde{x} = \pm\sigma\sqrt{ln\left( \frac{(N_{1}+N_{2})}{N_{1}-N_{2}(\frac{1+3\gamma}{1-\gamma})}\right)}
\end{equation}
Here, $N_{1}$ and $N_{2}$ represent singles and coincidence counts, respectively. The optical delay introduced between the signal and idler photons can be estimated using Eq. (\ref{eq:9}) with the proper knowledge of $N_{1}$ and $N_{2}$. To perform the FI-based measurements on the small optical delay between the paired photons, we first calibrated the 1 mm long PPKTP crystal-based HOM interferometer to estimate the parameters, $\alpha$, $\gamma$, and $\sigma$. We performed 16 scans of the HOM dip to ascertain the precise values of these defining parameters of the HOM interferometer.  We tried to resolve between two selected points on the HOM interference curve having the positive optical delays of, say, $x_{1}$ and $x_{2}$ with a separation of $\sim$5.51 nm apart. The optical delay corresponding to the maximum peak value of FI lies between points $x_{1}$ and $x_{2}$. We set the voltage to the PEA according to the calibration data shown in Fig. \ref{piezo calibration} to make the back-and-forth movement between the positions $x_{1}$ and $x_{2}$ and recorded the singles count, $N_{1}$, and coincidence counts, $N_{2}$ at both points. Although the exposure time for this iterative measurement was set to 50 ms, the electronic response and delay of the Piezo controller and TDC restricted the effective data acquisition frequency to $\sim$6 Hz.
Using all the experimental parameters in Eq. (\ref{eq:9}), we have estimated the values of $x_{1}$ and $x_{2}$ with the results shown in Fig. \ref{est delay}. 
\begin{figure}[h]
\centering
\includegraphics[width=\linewidth]{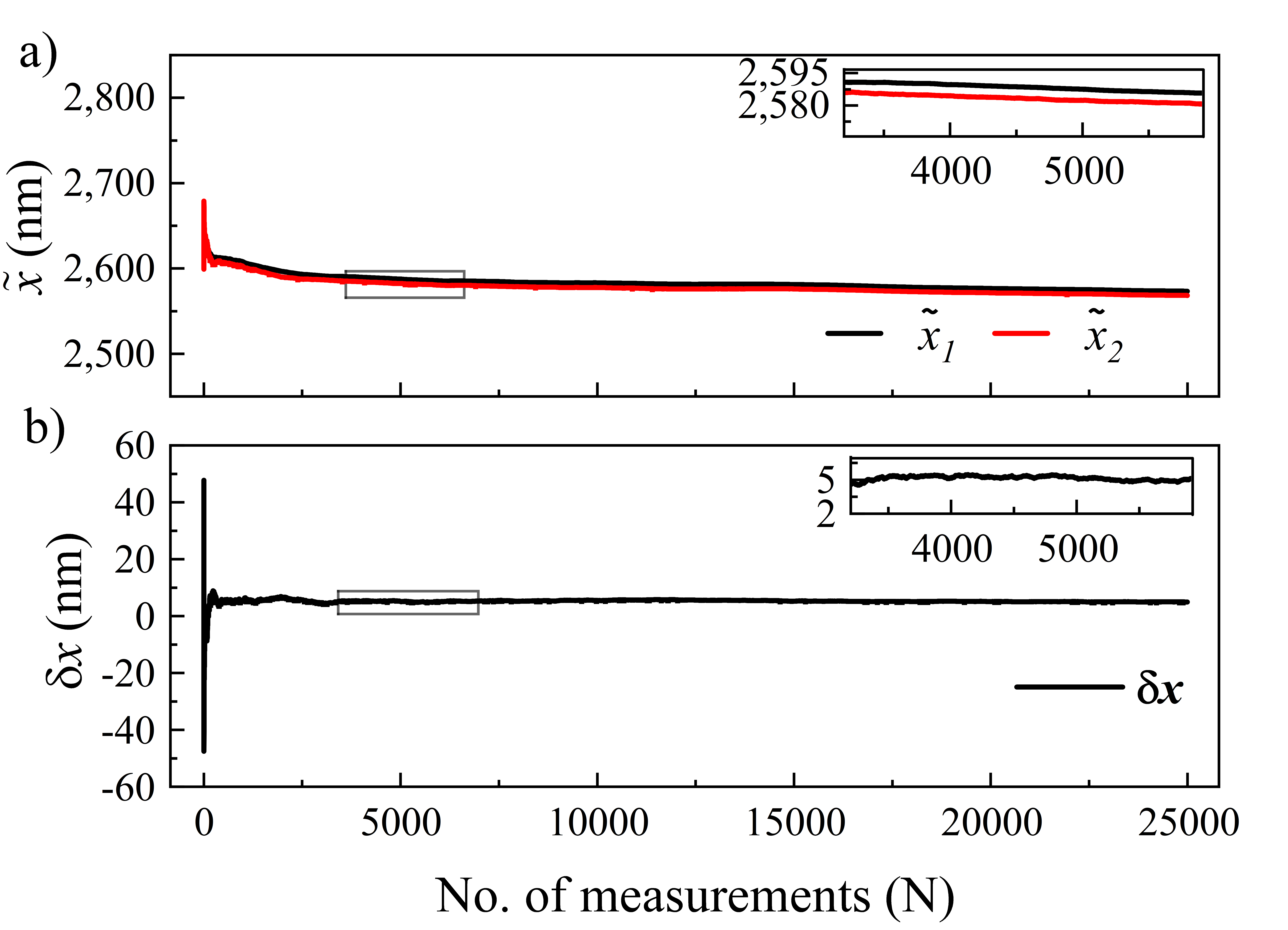}
\caption{Variation of a. cumulative estimates for the PZT positions,  $x_{1}$ and $x_{2}$, and b. the estimated optical delay between signal and idler photons as a function of the number of measurements.}
\label{est delay}
\end{figure}
As evident from Fig. \ref{est delay}(a), the cumulative estimates for $x_{1}$ (black line) and $x_{2}$ (red line) with the increase in the number of iterations/measurements show a drift due to the drift in the piezo position over the measurement time. However, the separation between the estimated values of $x_{1}$ and $x_{2}$, as shown by the inset image of Fig. \ref{est delay}(a), remains almost constant. We have estimated the separation between the set points as a function of the number of iterations with the results shown in Fig. \ref{est delay}(b). As evident from Fig. \ref{est delay}(b), we observe a large uncertainty in estimating the separation between the set positions for the initial measurements, which settles quickly toward the set value with the increase in the number of measurements/iterations. For the set separation value of $\delta x$ $\sim$ 5.51$\pm$0.75 nm estimated from the difference in the applied voltage to the PEA, we measured the optical delay to be 4.97 $\pm$ 0.89 nm even for the experimental measurements/iterations as low as 3300 (as shown in the inset of Fig. \ref{est delay}(b)). The achievable precision for the complete range of the measurements can be calculated as $\sqrt{Var(\delta \tilde{x})/N}$, where N is the total number of independent measurements performed for the cumulative estimation.\\
\indent
The minute drift in the cumulative estimated data (see from Fig. \ref{est delay}(a)), due to the drift in the piezo position over the large experiment time (as high as $\sim$ 139 minutes), can easily be mitigated by adjusting the frequency of back-and-forth movement of $\sim$ 6 Hz; a rate much faster than the rate of drift in the position of PEA. Despite such drift, it is to be noted that the PEA position remains well inside the region ($\sim$ 200 nm), having maximum FI value, adding to the reliability of the acquired data. Although we have performed about 25000 independent measurements over an experiment time of $\sim$139 minutes, one can easily notice from Fig. \ref{est delay}(b) that the true value has been achieved through the saturation of Cramér-Rao bound for the experimental measurements/iterations as low as 3380 corresponding to the experiment time as low as $\sim$19 minutes. This was possible due to the increase of the peak FI value through manipulating the spectral bandwidth of the SPDC photons. Further reduction in the experiment time to reach such high sensitivity can be possible by further enhancing the FI using the ultrafast pump laser generating SPDC photons in thin crystals.\\
\indent 
We repeated these measurements over a wide range of optical delays using PPKTP crystals of lengths 1, 2, 10, 20, and 30 mm. The estimated optical delays calculated  from the cumulative estimates are shown in Fig. \ref{est delay crystal lentghs}.  As evident from Fig. \ref{est delay crystal lentghs}, the experimental values of the cumulative estimates (solid dots) of the optical delay in the range of $\sim$3 nm to $\sim$80 nm measured using different crystal lengths exactly follow the set true value of $\delta x$ (black line) derived from the peak-to-peak amplitude of the voltage signal applied for the back-and-forth motion of the PEA. It is worth mentioning that we have incorporated the piezo hysteresis as shown in Fig. \ref{dynamic sensing} (b) in all the measurement data. The close agreement of the experimental values of the cumulative estimate with the set true values of $\delta x$, for all crystal lengths confirms the reliability and robustness of the current experimental scheme.\\
\begin{figure}[h]
\centering
\includegraphics[width=\linewidth]{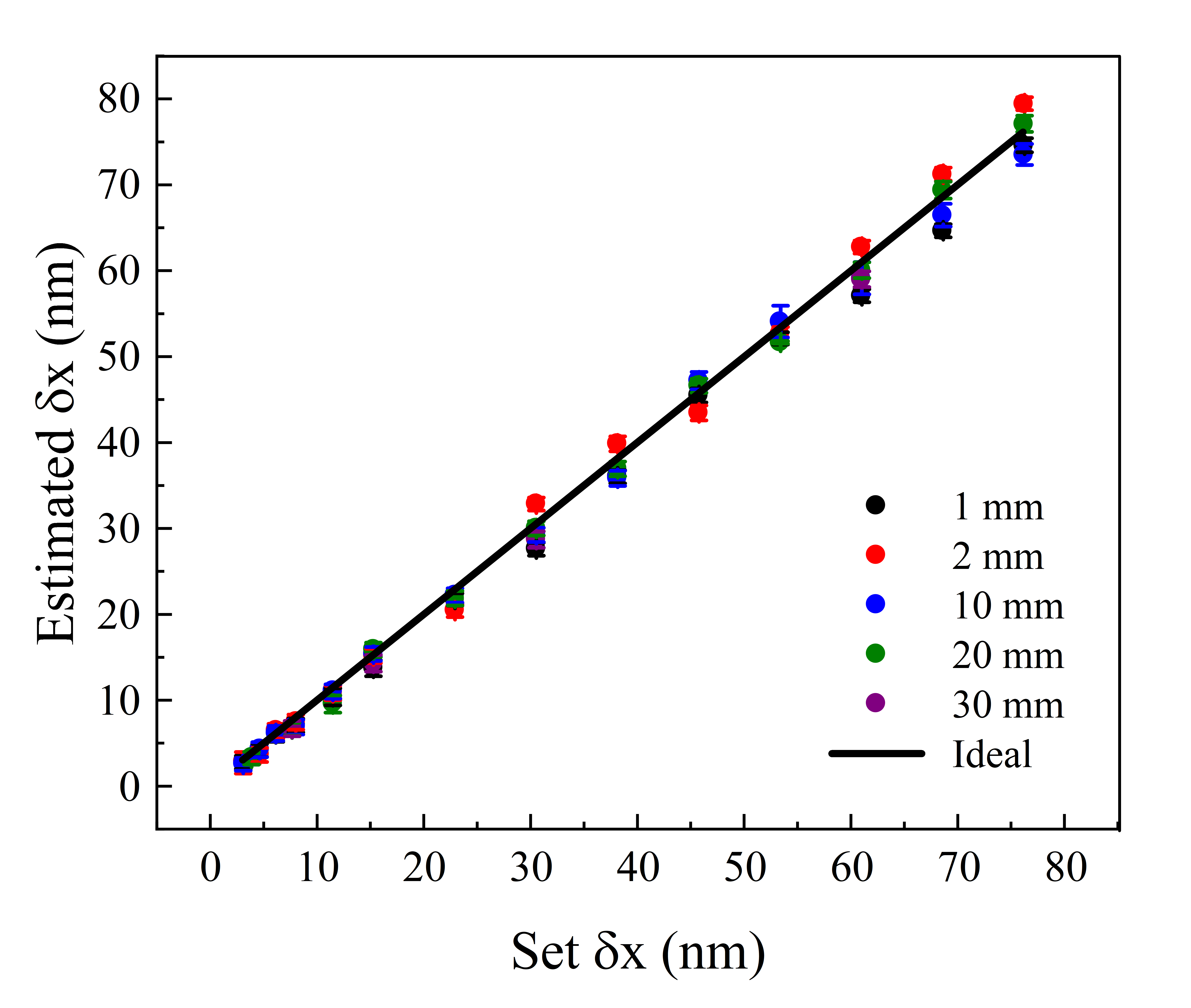}
\caption{Variation of optical delay measured using the FI analysis with the set optical delays for different crystal lengths. Solid line represents the true values.}
\label{est delay crystal lentghs}
\end{figure}
\indent
We further experimentally verify the dependence of the value of FI and the minimum number of required iterations/measurements while achieving a fixed precision. It is evident from Eq. (\ref{eq:6}) and Eq. (\ref{eq:8}) that for a given precision, the increase in the value of peak FI reduces the required number of iterations/measurements. For experimental verification, we set, for example, the precision of the optical delay between the pair photons to be $\sim$5 nm and observed the minimum number of measurements/iterations (N) required to achieve this precision for all available PPKTP crystals. The results are shown in Fig. \ref{Runs vs crystal length}. For better understanding, we 
%
\begin{figure}[h]
\centering
\includegraphics[width=\linewidth]{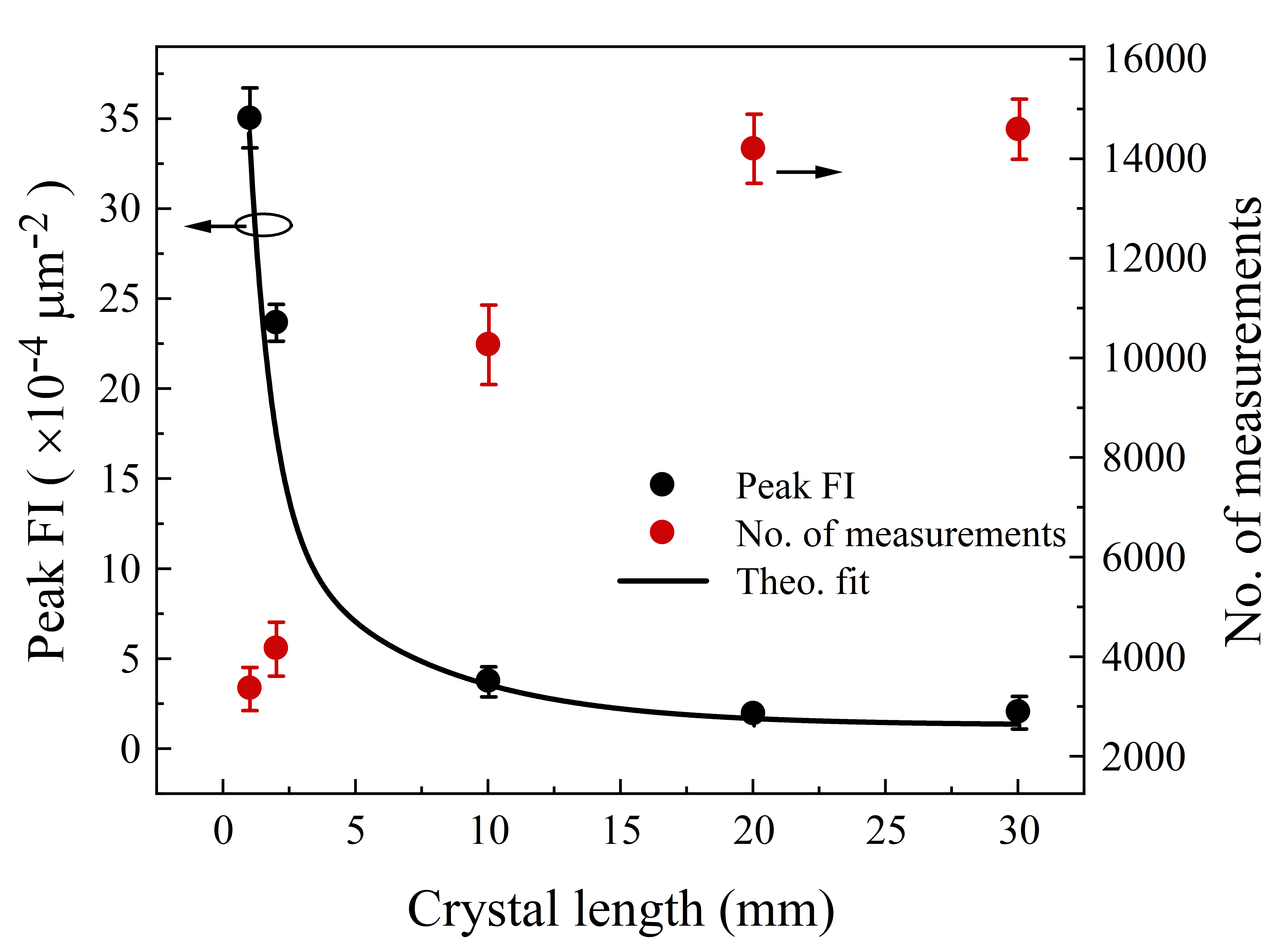}
\caption{Variation of peak FI value and the corresponding number of iterations/measurement required to achieve a fixed precision as a function of crystal length.}
\label{Runs vs crystal length}
\end{figure}
%
have also shown the variation of the peak value of FI with crystal length. As evident from Fig. \ref{Runs vs crystal length}, to achieve a set precision, the number of minimum iterations/measurements (red dots) increases from 3380 to 17500 for the increase of crystal length from 1 mm to 30 mm due to the subsequent decrease of peak FI value (black dots) from $\sim$35 x $10\textsuperscript{-4}$ to $\sim$2 x $10\textsuperscript{-4}$ $\upmu$m$\textsuperscript{-2}$. It is interesting to note that the inverse of the product of the number of minimum iterations/measurements (N) and the square root of the peak value of FI forming the precision \cite{chen2019biphotonbeat} is exactly matching with set precision and independent to the increase of crystal length from 1 mm to 30 mm. The black line is the theoretical fit to the experimental data as reproduced from Fig. \ref{FI vs crystal length}. From this experiment, it is evident that a further increase of the peak value of FI through the increase of spectral bandwidth of the SPDC photon using of ultrafast pump can lead to the realization of FI-based ultra-sensitive measurements in real-time applications.\\ 

\section*{Conclusions}
\noindent
In conclusion, we have experimentally demonstrated easy control in the spectral bandwidth of the pair-photons through proper selection of the length of the non-linear crystal. Using a 1 mm long PPKTP crystal, we have generated paired photons with spectral width as high as 163.42$\pm$1.68 nm even in the presence of a single-frequency, CW, diode laser as the pump. The use of photon pairs with such a high spectral bandwidth in a HOM interferometer results in a narrow-width HOM interference dip enabling sensing of static displacement as low as 60 nm and threshold vibration amplitude as low as $\sim$205 nm with a resolution of $\sim$80 nm at a frequency measurement up to 8 Hz. Further, we experimentally observed the dependence of FI on the spectral bandwidth of the pair photons and hence the length of the nonlinear crystal. We observed a 17 times enhancement in the FI value while reducing the crystal length from 30 mm to 1 mm. Such an increase in the  peak FI value (35.06 $\pm$ 1.65) x $10\textsuperscript{-4}$ $\upmu$m$\textsuperscript{-2}$, which is nearly 24 times higher than the previous study \cite{lyons2018attosecond}, saturates the Cramér-Rao bound to achieve any arbitrary precision (say $\sim$5 nm) in a lower number of  iterations ($\sim$ 3300), $\sim$11 times lower than the previous reports. Multiplying the number of coincidence counts to the number of iterations, one can find the total number of photons used in the experiment to achieve such high precision. In our case, this provides 2 $\times$ $10^6$ number of photon pairs to achieve the desired precision. This is somewhat higher than the 2 $\times$ $10^4$ photons used e.g. in the experiments by Kwait et.al.\cite{kwait2023}. However, our total measurement time is still lower thanks to the high photon flux rates, therefore allowing us to achieve overall faster sampling rates and sensitivity to higher vibration frequencies.   
The accessibility of high precision in lower iterations or time establishes the potential of HOM-based sensors for real-time, precision-augmented, in-field quantum sensing applications. Unlike the use of bulky and expensive ultrafast pump lasers of broad spectral bandwidth to enhance the spectral bandwidth of the down-converted photons, the generation of broadband photons using a single-frequency diode laser in small crystal length is beneficial for any practical applications. \\


\section*{Materials and Methods}\label{Methods}
\noindent
The schematic of the experimental configuration is shown in Fig. \ref{Exp}. A single-frequency, CW diode laser providing 20 mW of output power at 405.4 nm is used as the pump laser in the experiment. The spatial mode filtering unit comprised of a pair of fiber couplers (C1 and C2) and a single-mode fiber (SMF1) is used to transform the laser output in a Gaussian (TEM$_{00}$ mode) beam profile. The laser power in the experiment is controlled using a combination of $\lambda$/2 plate (HWP1) and a polarizing beam splitter cube (PBS1). The second $\lambda$/2 plate (HWP2) is used to adjust the direction of linear polarization of the pump laser with respect to the orientation of the crystal poling direction to ensure optimum phase matching conditions.The pump laser is focused at the center of the nonlinear crystal (NC) using the plano-convex lens, L1, of focal length, $f1$ = 75 mm to a beam waist radius of $w_{o}$ = 58 $\upmu$m. Periodically poled KTiOPO$_{4}$ (PPKTP) crystals of 1 $\times$ 2 mm$^2$ aperture but of five different interaction lengths, $L$ = 1, 2, 10, 20, and 30 mm, are used as the nonlinear crystal (NC) in the experiment. The crystals have a single grating period of $\Lambda$ = 3.425 $\upmu$m corresponding to the degenerate, type-0 ($e $→$ e + e$) phase-matched parametric down-conversion (PDC) of 405 nm at 810 nm. To ensure quasi-phase-matching, the PPKTP crystals are housed in an oven whose temperature can be varied from room temperature to 200$^o$C with temperature stability of $\pm$0.1$^o$C.
\noindent
The down-converted photon pairs, after separation from the pump photons using a notch filter (NF),  are collimated using the plano-convex lens, L2 of focal length, $f2$ = 100 mm. Due to the non-collinear phase-matching, the correlated pair (commonly identified as ``signal" and ``idler") of down-converted photons have annular ring spatial distribution with signal and idler photons on diametrically opposite points. Using a prism mirror (PM), the signal and idler photons are separated and guided in two different paths and finally made to interact on the two separate input ports of a balanced beam splitter (BS). Since we need to vary temporal delay between the photons at the BS, and the translation of the optical component disturbs the alignment, we have devised the alignment preserving delay path consisting of a mirror, M1, $\lambda/2$ plate (HWP3), PBS2, $\lambda/4$ plate (QWP), mirror, M2 on the delay stage, and finally, the beam splitter, BS. The $\lambda/2$ plate rotates the photon's polarization state from vertical to horizontal to ensure complete transmission through PBS2. The $\lambda/4$ having the fast axis at +45$^o$ to its polarization axis transforms the photon polarization (horizontal) into circular (left circular). However, the handedness of the photon polarization  gets reversed (right circular) on reflection from the mirror, M2, at normal incidence. On the return pass through the $\lambda/4$ plate, the right circular polarized photons convert into vertical polarized photons and are subsequently reflected from the PBS2 to one of the input ports of the BS. Due to the normal incident of the photons on the moving mirror, M2, the varying delay of the photons does not require tweaking of the experiment. The combination of the piezo-electric actuator (PEA) (NF15AP25), placed on a motorized linear translation stage (MTS25-Z8), is used to provide both fine and coarse optical delay. The range (resolution) of the PEA and motorized linear stage as specified by the product catalog is 25 $\upmu$m (0.75 nm), and 25 mm (29 nm), respectively. Throughout the report, the optical delay of the photon is twice the displacement of the mirror, M2, if otherwise mentioned. On the other hand, the photons of the other arm are guided using mirrors, M3 and M4, to the second input port of BS such that the optical path length of both arms are the same. The photons from the output ports of the BS, after being extracted by long pass filters (F1, F2), are coupled into the single mode fibers, SMF2 and SMF3, through the fiber couplers, C3, and C4, respectively. The collected photons are detected using the single-photon counting modules, SPCM1-2 (AQRH-14-FC, Excelitas), and counted using the time-to-digital converter (TDC). All optical components used in the experiment are selected for optimum performance at both the pump and SPDC wavelengths. All the data is recorded at a typical pump power of 0.25 mW with a coincidence window of 1.6 nanoseconds.\\
\subsection*{Characterization of HOM interference}\label{results}
\noindent
First, we verified the dependence of HOM characteristics on the length of the nonlinear crystal. Keeping the input pump power to the crystal and the temporal coincidence window of the TDC constant at 0.25 mW and 1.6 ns, respectively, we have recorded the variation of coincidence counts as a function of the relative optical delay between the photons. As evident from Fig. \ref{crystal len vs spectral bandwidth}(a), the variation of coincidence counts with the relative optical path delay has the characteristic HOM interference dip profile, approximated by an inverted Gaussian function (owing to the temporal Gaussian profile of signal and idler). However, it is interesting to note that the HOM visibilities calculated using the formula given in \cite{lyons2018attosecond} remain constant in the range of 86 - 94 \% despite appreciable variation in the span of the HOM interference dip profile. \\
\begin{figure}[h]
\centering
\includegraphics[width=\linewidth]{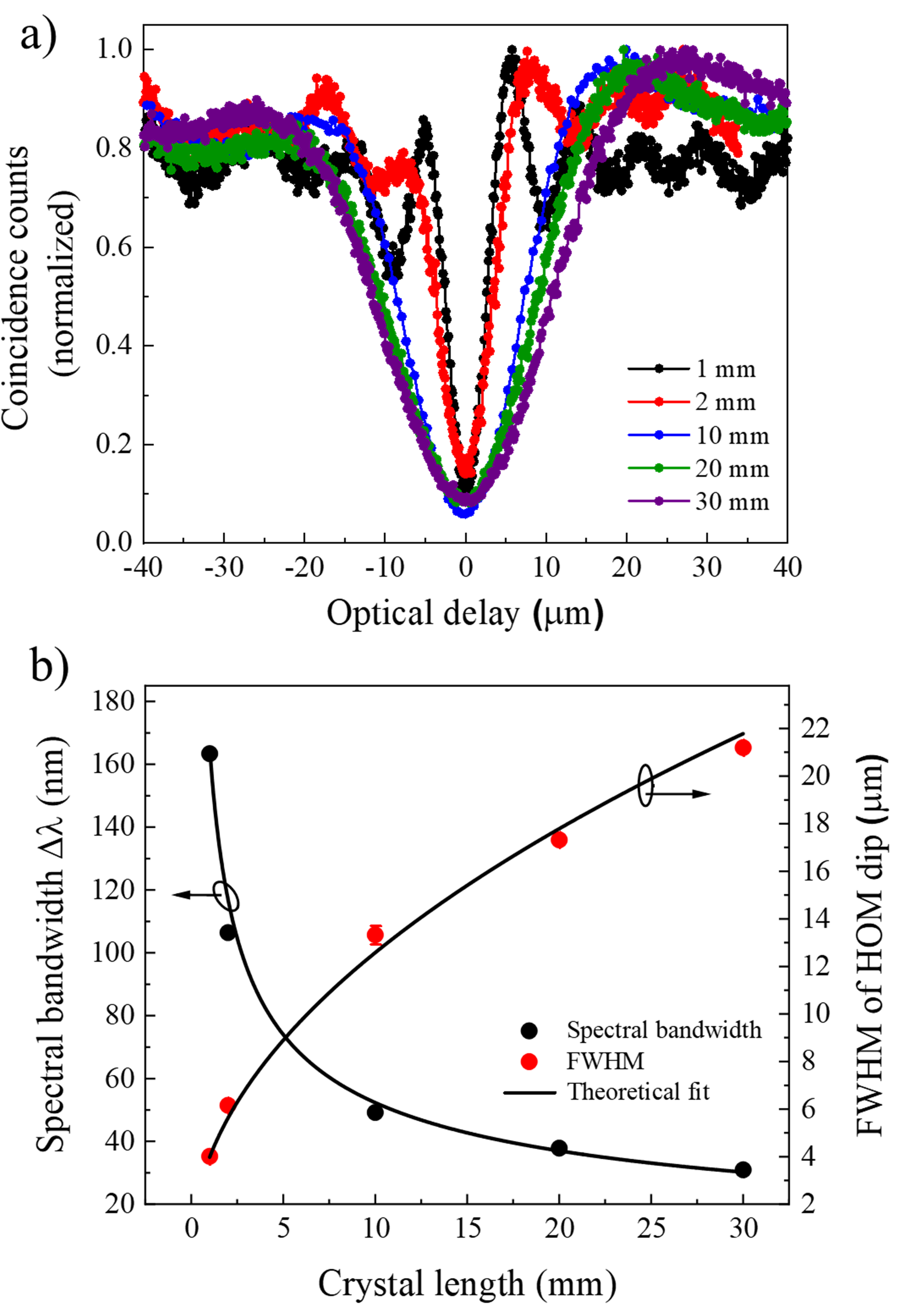}
\caption{a. Variation of HOM interference dip profiles with the length of the non-linear crystals. b. Dependence of spectral bandwidth (black dots) and the FWHM width of HOM interference profile (red dots) of SPDC photons on the length of the non-linear crystal. The solid black line is theoretical fit to the experimental data.}
\label{crystal len vs spectral bandwidth}
\end{figure}
\indent
To gain further insights, we measured the FWHM width of the HOM interference dip and estimated the corresponding spectral bandwidth of the {photon pairs using Eq. (\ref{equ:4}) and (\ref{equ:5}). The results are shown in Fig. \ref{crystal len vs spectral bandwidth}(b). It is evident from Fig. \ref{crystal len vs spectral bandwidth}(b) that the HOM interference dip has FWHM width (red dots) of 4.01 $\pm$ 0.04, 6.16 $\pm$ 0.05, 13.33 $\pm$ 0.40, 17.32 $\pm$ 0.08 and 21.19 $\pm$ 0.13 $\upmu$m corresponding to the estimated FWHM spectral bandwidth (black dots) of SPDC photons of 163 $\pm$ 1.68, 106 $\pm$ 0.93, 49 $\pm$ 1.44, 38 $\pm$ 0.20 and 30 $\pm$ 0.18 nm for the crystal lengths of 1, 2, 5, 10, 20, and 30 mm are in close agreement with the Eq. \ref{equ:5} (solid black lines). Using the available pump power, we experimentally measured the spectral bandwidth of the SPDC photons using the spectrometer (HR-4000, Ocean Optics) to be $\sim$30 nm, same as our previous report \cite{Jabir:17} and $\sim$37 nm for 30 mm and 20 mm long crystals, respectively. The relatively low parametric gain has restricted the experimental measurement of the SPDC spectrum for smaller crystal lengths. However, it is important to note that the experimentally measured spectral bandwidth matches the estimated spectral bandwidth as shown in Fig. \ref{crystal len vs spectral bandwidth}(b). Interestingly, the decrease of crystal length from 30 mm to 1 mm results in a decrease (increase) of HOM interference dip width (spectral width of SPDC photons) by more than five times. As a result, one can see the sharp variation (see black and red dots of Fig. \ref{crystal len vs spectral bandwidth}(a)) in the coincidence counts for the small changes in the optical delay away from  the zero-point optical delay for HOM interference with smaller dip width. Such amplification in the rate of change of coincidence counts with optical delay, i.e., the increase of sensitivity of the HOM interferometer with the decrease in crystal length, can be useful for designing HOM interferometer-based sensor to measure any physical process influencing the optical delay between the photons.\\
\indent
To gain further perspective, we have selected half of the HOM curve of Fig. \ref{crystal len vs spectral bandwidth}(a) showing coincidence variation from minimum to maximum with positive optical delay for crystal lengths, $L$ = 1, 2, 10, 20 and 30 mm. The results are shown in Fig. \ref{HOM curve slopes}. Since the maximum coincidence counts per second at a fixed pump power of the HOM curve directly depend on the pair generation rate and hence the interaction length of the non-linear crystal used, for comparative study of the slopes of HOM curves for different crystal lengths, we have normalized the coincidence counts with its maximum. Using the linear fit (solid lines) to the experimental data (points), we observe
\begin{figure}[h]
\centering
\includegraphics[width=\linewidth]{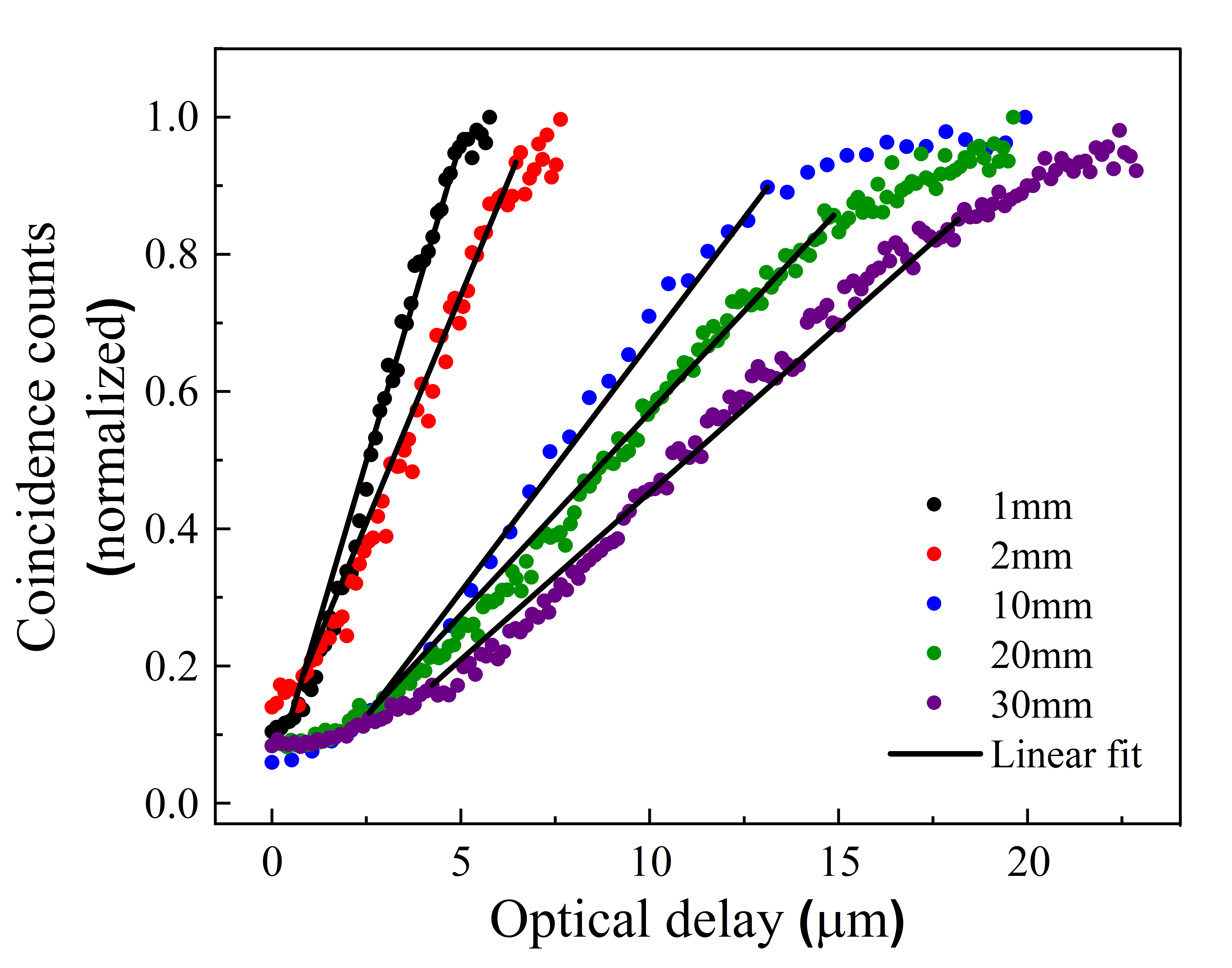}
\caption{Crystal length dependent variation in the slope of coincidence counts of HOM interferometer. Solid lines are linear fit to the experimental results (dots).}
\label{HOM curve slopes}
\end{figure}
the slope (see Fig. \ref{HOM curve slopes}), of the HOM curve to decrease with the crystal length from the maximum of 0.215 /$\upmu$m for $L$ = 1 mm to 0.051 /$\upmu$m for $L$ = 30 mm. In terms of the coincidence counts, the slopes of the HOM curve for 1 mm and 30 mm crystals are found to be 450/$\upmu$m and 260/$\upmu$m, respectively, at a pump power of 0.25 mw and exposure time at 20 ms. As expected, due to the high slope or sensitivity of the HOM interferometer, the range is low for $L$ = 1 mm than for $L$ = 30 mm. However, it is evident that in order to have access to the high sensitivity of the HOM interferometer for the measurement of any physical quantity, it is advisable to use a shorter nonlinear crystal length. However, one has to be careful with the overall coincidence counts arising from the lower parametric gain due to the small crystal length. On the other hand, in the present study, we have used a CW diode laser to implement the system in outdoor applications. However, using an ultrafast pump laser in combination with the thin crystal for lab-based experiments, one can further enhance the sensitivity of the HOM interferometer due to the increase of overall spectral bandwidth (decrease of HOM interference dip width) of the down-converted photons. 
\subsection*{Calibration of piezo actuator}\label{calibration}
%
\begin{figure}[h]
\centering
\includegraphics[width=\linewidth]{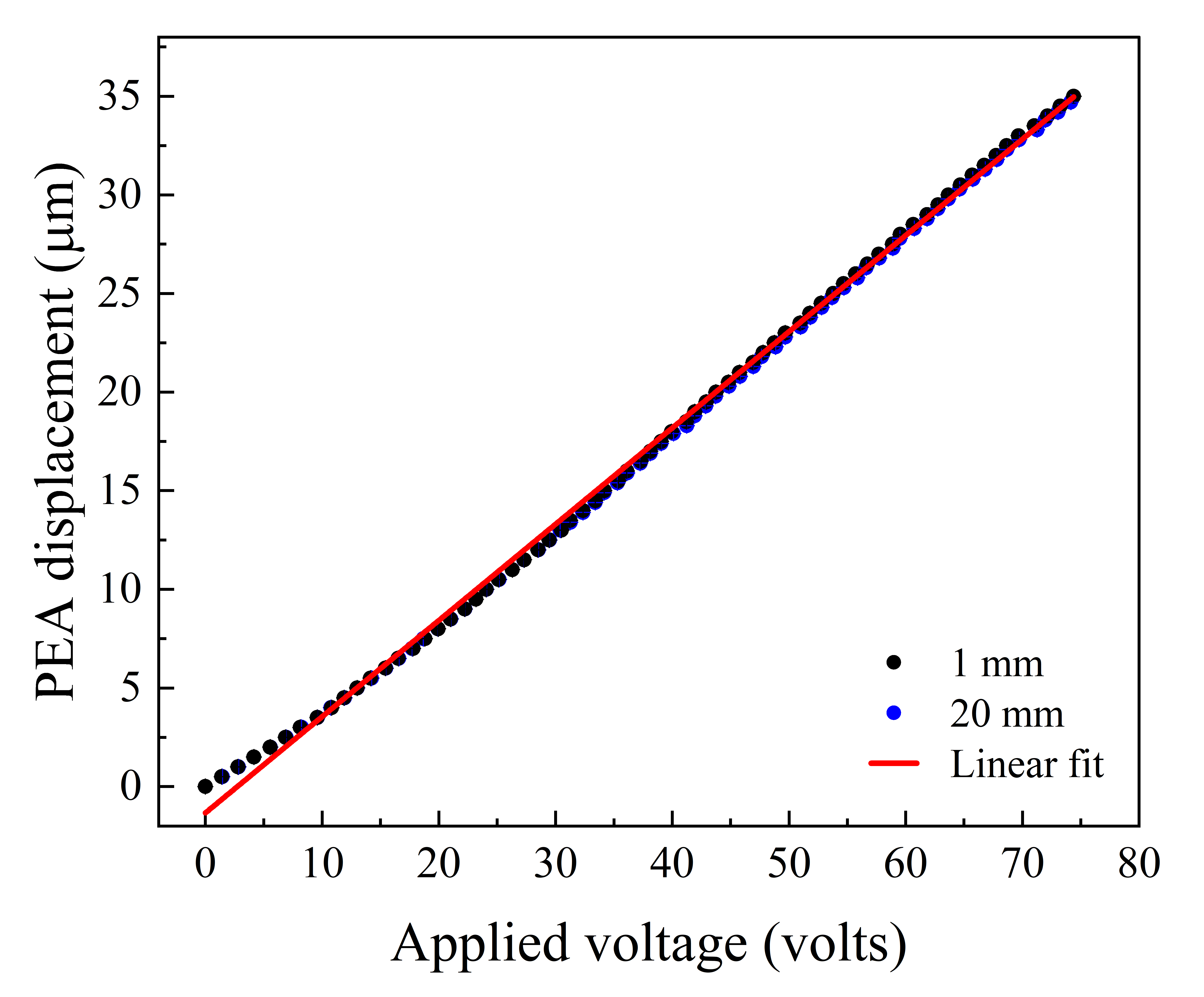}
\caption{Linear displacement of the PEA stage as a function of applied voltage.}
\label{piezo calibration}
\end{figure}
\noindent
After successful characterization and static displacement measurement, we used a piezo-electric actuator (PEA) (NF15AP25), having a travel range of 25 $\upmu$m for the applied voltage 0 - 75 V to study the performance of the HOM interferometer for dynamic optical delay or mirror vibration sensing applications. However, due to the unavailability of closed-loop feedback for the PEA, we had to calibrate the PEA to estimate the displacement with the applied voltage. In doing so, we have attached the mirror M2 (see Fig. \ref{Exp}) to PEA placed over the motorized linear translation stage (MTS25-Z8). Using a crystal of length 20 mm corresponding to the HOM curve represented by the green dots in Fig. \ref{crystal len vs spectral bandwidth}, we have adjusted the initial position of the linear stage corresponding to a positive delay of $\sim$10 $\upmu$m so that the coincidence count is 50\% of maximum value. We moved the linear stage by $\sim$0.5 $\upmu$m corresponding to the optical delay of $\sim$1 $\upmu$m towards the zero-delay point resulting in a change in the coincidence counts. Such change in the coincidence counts is reset by increasing the applied voltage to the PEA. Under this condition, the applied voltage to the PEA makes a displacement corresponding to the displacement of the linear stage in the opposite direction. We have repeated this exercise in steps of $\sim$0.5 $\upmu$m up to the allowed PEA voltage of 75 V with the results shown in Fig. \ref{piezo calibration}. It is evident from Fig. \ref{piezo calibration} that the displacement (blue dots) of the PEA is linear to the applied voltage producing a maximum displacement of 35 $\upmu$m, more than the manufacturer's specification (25 $\upmu$m) for the maximum allowed voltage of 75 V. Using the linear fit (red line) to the experimental data (black dots) we find the displacement of the PEA to be 0.48 $\pm$ 0.02 $\upmu$m per volt. As expected, the calibration of PEA using the HOM interferometer based on a 1 mm long crystal (black dots) coincides with the results obtained using the 20 mm long crystal (blue dots).

\vspace{12pt}
\indent
\textbf{Acknowledgments} We acknowledge the support of Ms. Shreya Mishra, Physical Research Laboratory, Ahmedabad, for the useful discussion on statistics. \par
%
\indent
\textbf{Conflict of Interest:} All authors declare no financial/commercial conflicts of interest. \par
%
%
\indent
\textbf{Data Availability Statement:} The data that support the findings of this study are available from the corresponding author upon reasonable request.}\par

\bibliography{reference}

\begin{thebibliography}{41}%
\makeatletter
\providecommand \@ifxundefined [1]{%
 \@ifx{#1\undefined}
}%
\providecommand \@ifnum [1]{%
 \ifnum #1\expandafter \@firstoftwo
 \else \expandafter \@secondoftwo
 \fi
}%
\providecommand \@ifx [1]{%
 \ifx #1\expandafter \@firstoftwo
 \else \expandafter \@secondoftwo
 \fi
}%
\providecommand \natexlab [1]{#1}%
\providecommand \enquote  [1]{``#1''}%
\providecommand \bibnamefont  [1]{#1}%
\providecommand \bibfnamefont [1]{#1}%
\providecommand \citenamefont [1]{#1}%
\providecommand \href@noop [0]{\@secondoftwo}%
\providecommand \href [0]{\begingroup \@sanitize@url \@href}%
\providecommand \@href[1]{\@@startlink{#1}\@@href}%
\providecommand \@@href[1]{\endgroup#1\@@endlink}%
\providecommand \@sanitize@url [0]{\catcode `\\12\catcode `\$12\catcode
  `\&12\catcode `\#12\catcode `\^12\catcode `\_12\catcode `\%12\relax}%
\providecommand \@@startlink[1]{}%
\providecommand \@@endlink[0]{}%
\providecommand \url  [0]{\begingroup\@sanitize@url \@url }%
\providecommand \@url [1]{\endgroup\@href {#1}{\urlprefix }}%
\providecommand \urlprefix  [0]{URL }%
\providecommand \Eprint [0]{\href }%
\providecommand \doibase [0]{https://doi.org/}%
\providecommand \selectlanguage [0]{\@gobble}%
\providecommand \bibinfo  [0]{\@secondoftwo}%
\providecommand \bibfield  [0]{\@secondoftwo}%
\providecommand \translation [1]{[#1]}%
\providecommand \BibitemOpen [0]{}%
\providecommand \bibitemStop [0]{}%
\providecommand \bibitemNoStop [0]{.\EOS\space}%
\providecommand \EOS [0]{\spacefactor3000\relax}%
\providecommand \BibitemShut  [1]{\csname bibitem#1\endcsname}%
\let\auto@bib@innerbib\@empty
\bibitem [{\citenamefont {Pirandola}\ \emph {et~al.}(2018)\citenamefont
  {Pirandola}, \citenamefont {Bardhan}, \citenamefont {Gehring}, \citenamefont
  {Weedbrook},\ and\ \citenamefont {Lloyd}}]{2018QS}%
  \BibitemOpen
  \bibfield  {author} {\bibinfo {author} {\bibfnamefont {S.}~\bibnamefont
  {Pirandola}}, \bibinfo {author} {\bibfnamefont {B.~R.}\ \bibnamefont
  {Bardhan}}, \bibinfo {author} {\bibfnamefont {T.}~\bibnamefont {Gehring}},
  \bibinfo {author} {\bibfnamefont {C.}~\bibnamefont {Weedbrook}},\ and\
  \bibinfo {author} {\bibfnamefont {S.}~\bibnamefont {Lloyd}},\ }\bibfield
  {title} {\bibinfo {title} {Advances in photonic quantum sensing},\
  }\href@noop {} {\bibfield  {journal} {\bibinfo  {journal} {Nature Photonics}\
  }\textbf {\bibinfo {volume} {12}},\ \bibinfo {pages} {724} (\bibinfo {year}
  {2018})}\BibitemShut {NoStop}%
\bibitem [{\citenamefont {Bian}\ \emph {et~al.}(2021)\citenamefont {Bian},
  \citenamefont {Zheng}, \citenamefont {Zeng}, \citenamefont {Chen},
  \citenamefont {St{\"o}hr}, \citenamefont {Denisenko}, \citenamefont {Yang},
  \citenamefont {Wrachtrup},\ and\ \citenamefont
  {Jiang}}]{bian2021electricnanoscale}%
  \BibitemOpen
  \bibfield  {author} {\bibinfo {author} {\bibfnamefont {K.}~\bibnamefont
  {Bian}}, \bibinfo {author} {\bibfnamefont {W.}~\bibnamefont {Zheng}},
  \bibinfo {author} {\bibfnamefont {X.}~\bibnamefont {Zeng}}, \bibinfo {author}
  {\bibfnamefont {X.}~\bibnamefont {Chen}}, \bibinfo {author} {\bibfnamefont
  {R.}~\bibnamefont {St{\"o}hr}}, \bibinfo {author} {\bibfnamefont
  {A.}~\bibnamefont {Denisenko}}, \bibinfo {author} {\bibfnamefont
  {S.}~\bibnamefont {Yang}}, \bibinfo {author} {\bibfnamefont {J.}~\bibnamefont
  {Wrachtrup}},\ and\ \bibinfo {author} {\bibfnamefont {Y.}~\bibnamefont
  {Jiang}},\ }\bibfield  {title} {\bibinfo {title} {Nanoscale electric-field
  imaging based on a quantum sensor and its charge-state control under ambient
  condition},\ }\href@noop {} {\bibfield  {journal} {\bibinfo  {journal}
  {Nature Communications}\ }\textbf {\bibinfo {volume} {12}},\ \bibinfo {pages}
  {2457} (\bibinfo {year} {2021})}\BibitemShut {NoStop}%
\bibitem [{\citenamefont {Tsukamoto}\ \emph {et~al.}(2022)\citenamefont
  {Tsukamoto}, \citenamefont {Ito}, \citenamefont {Ogawa}, \citenamefont
  {Ashida}, \citenamefont {Sasaki},\ and\ \citenamefont
  {Kobayashi}}]{tsukamoto2022magneticaccurate}%
  \BibitemOpen
  \bibfield  {author} {\bibinfo {author} {\bibfnamefont {M.}~\bibnamefont
  {Tsukamoto}}, \bibinfo {author} {\bibfnamefont {S.}~\bibnamefont {Ito}},
  \bibinfo {author} {\bibfnamefont {K.}~\bibnamefont {Ogawa}}, \bibinfo
  {author} {\bibfnamefont {Y.}~\bibnamefont {Ashida}}, \bibinfo {author}
  {\bibfnamefont {K.}~\bibnamefont {Sasaki}},\ and\ \bibinfo {author}
  {\bibfnamefont {K.}~\bibnamefont {Kobayashi}},\ }\bibfield  {title} {\bibinfo
  {title} {Accurate magnetic field imaging using nanodiamond quantum sensors
  enhanced by machine learning},\ }\href@noop {} {\bibfield  {journal}
  {\bibinfo  {journal} {Scientific Reports}\ }\textbf {\bibinfo {volume}
  {12}},\ \bibinfo {pages} {13942} (\bibinfo {year} {2022})}\BibitemShut
  {NoStop}%
\bibitem [{\citenamefont {Hendricks}\ \emph {et~al.}(2018)\citenamefont
  {Hendricks}, \citenamefont {Ahmed}, \citenamefont {Barker}, \citenamefont
  {Egan}, \citenamefont {Douglass}, \citenamefont {Eckel}, \citenamefont
  {Fedchak}, \citenamefont {Klimov}, \citenamefont {Ricker}, \citenamefont
  {Scherschligt} \emph {et~al.}}]{NISTquantumsensor}%
  \BibitemOpen
  \bibfield  {author} {\bibinfo {author} {\bibfnamefont {J.}~\bibnamefont
  {Hendricks}}, \bibinfo {author} {\bibfnamefont {Z.}~\bibnamefont {Ahmed}},
  \bibinfo {author} {\bibfnamefont {D.}~\bibnamefont {Barker}}, \bibinfo
  {author} {\bibfnamefont {P.}~\bibnamefont {Egan}}, \bibinfo {author}
  {\bibfnamefont {K.}~\bibnamefont {Douglass}}, \bibinfo {author}
  {\bibfnamefont {S.}~\bibnamefont {Eckel}}, \bibinfo {author} {\bibfnamefont
  {J.}~\bibnamefont {Fedchak}}, \bibinfo {author} {\bibfnamefont
  {N.}~\bibnamefont {Klimov}}, \bibinfo {author} {\bibfnamefont
  {J.}~\bibnamefont {Ricker}}, \bibinfo {author} {\bibfnamefont
  {J.}~\bibnamefont {Scherschligt}}, \emph {et~al.},\ }\bibfield  {title}
  {\bibinfo {title} {The emerging field of quantum based measurements for
  pressure, vacuum and beyond},\ }\href@noop {} {\bibfield  {journal} {\bibinfo
   {journal} {journal of physics: conference proceedings}\ }\textbf {\bibinfo
  {volume} {1065}},\ \bibinfo {pages} {162017} (\bibinfo {year}
  {2018})}\BibitemShut {NoStop}%
\bibitem [{\citenamefont {Kakuyanagi}\ \emph {et~al.}(2023)\citenamefont
  {Kakuyanagi}, \citenamefont {Toida}, \citenamefont {Abdurakhimov},\ and\
  \citenamefont {Saito}}]{Tempsensor}%
  \BibitemOpen
  \bibfield  {author} {\bibinfo {author} {\bibfnamefont {K.}~\bibnamefont
  {Kakuyanagi}}, \bibinfo {author} {\bibfnamefont {H.}~\bibnamefont {Toida}},
  \bibinfo {author} {\bibfnamefont {L.~V.}\ \bibnamefont {Abdurakhimov}},\ and\
  \bibinfo {author} {\bibfnamefont {S.}~\bibnamefont {Saito}},\ }\bibfield
  {title} {\bibinfo {title} {Submicrometer-scale temperature sensing using
  quantum coherence of a superconducting qubit},\ }\href@noop {} {\bibfield
  {journal} {\bibinfo  {journal} {New Journal of Physics}\ } (\bibinfo {year}
  {2023})}\BibitemShut {NoStop}%
\bibitem [{\citenamefont {Shi}\ \emph {et~al.}(2020)\citenamefont {Shi},
  \citenamefont {Li}, \citenamefont {Chen}, \citenamefont {Qin}, \citenamefont
  {Jiang},\ and\ \citenamefont {Wu}}]{pressuresensor}%
  \BibitemOpen
  \bibfield  {author} {\bibinfo {author} {\bibfnamefont {L.}~\bibnamefont
  {Shi}}, \bibinfo {author} {\bibfnamefont {Z.}~\bibnamefont {Li}}, \bibinfo
  {author} {\bibfnamefont {M.}~\bibnamefont {Chen}}, \bibinfo {author}
  {\bibfnamefont {Y.}~\bibnamefont {Qin}}, \bibinfo {author} {\bibfnamefont
  {Y.}~\bibnamefont {Jiang}},\ and\ \bibinfo {author} {\bibfnamefont
  {L.}~\bibnamefont {Wu}},\ }\bibfield  {title} {\bibinfo {title} {Quantum
  effect-based flexible and transparent pessure sensors with ultrahigh
  sensitivity and sensing density},\ }\href@noop {} {\bibfield  {journal}
  {\bibinfo  {journal} {Nature Communications}\ }\textbf {\bibinfo {volume}
  {11}},\ \bibinfo {pages} {3529} (\bibinfo {year} {2020})}\BibitemShut
  {NoStop}%
\bibitem [{\citenamefont {Degen}\ \emph {et~al.}(2017)\citenamefont {Degen},
  \citenamefont {Reinhard},\ and\ \citenamefont
  {Cappellaro}}]{RevModPhys.89.035002}%
  \BibitemOpen
  \bibfield  {author} {\bibinfo {author} {\bibfnamefont {C.~L.}\ \bibnamefont
  {Degen}}, \bibinfo {author} {\bibfnamefont {F.}~\bibnamefont {Reinhard}},\
  and\ \bibinfo {author} {\bibfnamefont {P.}~\bibnamefont {Cappellaro}},\
  }\bibfield  {title} {\bibinfo {title} {Quantum sensing},\ }\href
  {https://doi.org/10.1103/RevModPhys.89.035002} {\bibfield  {journal}
  {\bibinfo  {journal} {Rev. Mod. Phys.}\ }\textbf {\bibinfo {volume} {89}},\
  \bibinfo {pages} {035002} (\bibinfo {year} {2017})}\BibitemShut {NoStop}%
\bibitem [{\citenamefont {Hong}\ \emph {et~al.}(1987)\citenamefont {Hong},
  \citenamefont {Ou},\ and\ \citenamefont {Mandel}}]{HOM.PhysRevLett.59.2044}%
  \BibitemOpen
  \bibfield  {author} {\bibinfo {author} {\bibfnamefont {C.~K.}\ \bibnamefont
  {Hong}}, \bibinfo {author} {\bibfnamefont {Z.~Y.}\ \bibnamefont {Ou}},\ and\
  \bibinfo {author} {\bibfnamefont {L.}~\bibnamefont {Mandel}},\ }\bibfield
  {title} {\bibinfo {title} {Measurement of subpicosecond time intervals
  between two photons by interference},\ }\href
  {https://doi.org/10.1103/PhysRevLett.59.2044} {\bibfield  {journal} {\bibinfo
   {journal} {Phys. Rev. Lett.}\ }\textbf {\bibinfo {volume} {59}},\ \bibinfo
  {pages} {2044} (\bibinfo {year} {1987})}\BibitemShut {NoStop}%
\bibitem [{\citenamefont {Steinberg}\ \emph {et~al.}(1992)\citenamefont
  {Steinberg}, \citenamefont {Kwiat},\ and\ \citenamefont
  {Chiao}}]{dispersioncancellation}%
  \BibitemOpen
  \bibfield  {author} {\bibinfo {author} {\bibfnamefont {A.~M.}\ \bibnamefont
  {Steinberg}}, \bibinfo {author} {\bibfnamefont {P.~G.}\ \bibnamefont
  {Kwiat}},\ and\ \bibinfo {author} {\bibfnamefont {R.~Y.}\ \bibnamefont
  {Chiao}},\ }\bibfield  {title} {\bibinfo {title} {Dispersion cancellation and
  high-resolution time measurements in a fourth-order optical interferometer},\
  }\href {https://doi.org/10.1103/PhysRevA.45.6659} {\bibfield  {journal}
  {\bibinfo  {journal} {Phys. Rev. A}\ }\textbf {\bibinfo {volume} {45}},\
  \bibinfo {pages} {6659} (\bibinfo {year} {1992})}\BibitemShut {NoStop}%
\bibitem [{\citenamefont {Cassemiro}\ \emph
  {et~al.}(2010{\natexlab{a}})\citenamefont {Cassemiro}, \citenamefont
  {Laiho},\ and\ \citenamefont {Silberhorn}}]{2010accessing}%
  \BibitemOpen
  \bibfield  {author} {\bibinfo {author} {\bibfnamefont {K.~N.}\ \bibnamefont
  {Cassemiro}}, \bibinfo {author} {\bibfnamefont {K.}~\bibnamefont {Laiho}},\
  and\ \bibinfo {author} {\bibfnamefont {C.}~\bibnamefont {Silberhorn}},\
  }\bibfield  {title} {\bibinfo {title} {Accessing the purity of a single
  photon by the width of the hong--ou--mandel interference},\ }\href@noop {}
  {\bibfield  {journal} {\bibinfo  {journal} {New Journal of Physics}\ }\textbf
  {\bibinfo {volume} {12}},\ \bibinfo {pages} {113052} (\bibinfo {year}
  {2010}{\natexlab{a}})}\BibitemShut {NoStop}%
\bibitem [{\citenamefont {Cassemiro}\ \emph
  {et~al.}(2010{\natexlab{b}})\citenamefont {Cassemiro}, \citenamefont
  {Laiho},\ and\ \citenamefont {Silberhorn}}]{Cassemiro2010}%
  \BibitemOpen
  \bibfield  {author} {\bibinfo {author} {\bibfnamefont {K.~N.}\ \bibnamefont
  {Cassemiro}}, \bibinfo {author} {\bibfnamefont {K.}~\bibnamefont {Laiho}},\
  and\ \bibinfo {author} {\bibfnamefont {C.}~\bibnamefont {Silberhorn}},\
  }\bibfield  {title} {\bibinfo {title} {Accessing the purity of a single
  photon by the width of the hong–ou–mandel interference},\ }\href
  {https://doi.org/10.1088/1367-2630/12/11/113052} {\bibfield  {journal}
  {\bibinfo  {journal} {New Journal of Physics}\ }\textbf {\bibinfo {volume}
  {12}},\ \bibinfo {pages} {113052} (\bibinfo {year}
  {2010}{\natexlab{b}})}\BibitemShut {NoStop}%
\bibitem [{\citenamefont {Lyons}\ \emph {et~al.}(2018)\citenamefont {Lyons},
  \citenamefont {Knee}, \citenamefont {Bolduc}, \citenamefont {Roger},
  \citenamefont {Leach}, \citenamefont {Gauger},\ and\ \citenamefont
  {Faccio}}]{lyons2018attosecond}%
  \BibitemOpen
  \bibfield  {author} {\bibinfo {author} {\bibfnamefont {A.}~\bibnamefont
  {Lyons}}, \bibinfo {author} {\bibfnamefont {G.~C.}\ \bibnamefont {Knee}},
  \bibinfo {author} {\bibfnamefont {E.}~\bibnamefont {Bolduc}}, \bibinfo
  {author} {\bibfnamefont {T.}~\bibnamefont {Roger}}, \bibinfo {author}
  {\bibfnamefont {J.}~\bibnamefont {Leach}}, \bibinfo {author} {\bibfnamefont
  {E.~M.}\ \bibnamefont {Gauger}},\ and\ \bibinfo {author} {\bibfnamefont
  {D.}~\bibnamefont {Faccio}},\ }\bibfield  {title} {\bibinfo {title}
  {Attosecond-resolution hong-ou-mandel interferometry},\ }\href@noop {}
  {\bibfield  {journal} {\bibinfo  {journal} {Science advances}\ }\textbf
  {\bibinfo {volume} {4}},\ \bibinfo {pages} {eaap9416} (\bibinfo {year}
  {2018})}\BibitemShut {NoStop}%
\bibitem [{\citenamefont {Chen}\ \emph
  {et~al.}(2019{\natexlab{a}})\citenamefont {Chen}, \citenamefont {Fink},
  \citenamefont {Steinlechner}, \citenamefont {Torres},\ and\ \citenamefont
  {Ursin}}]{chen2019biphotonbeat}%
  \BibitemOpen
  \bibfield  {author} {\bibinfo {author} {\bibfnamefont {Y.}~\bibnamefont
  {Chen}}, \bibinfo {author} {\bibfnamefont {M.}~\bibnamefont {Fink}}, \bibinfo
  {author} {\bibfnamefont {F.}~\bibnamefont {Steinlechner}}, \bibinfo {author}
  {\bibfnamefont {J.~P.}\ \bibnamefont {Torres}},\ and\ \bibinfo {author}
  {\bibfnamefont {R.}~\bibnamefont {Ursin}},\ }\bibfield  {title} {\bibinfo
  {title} {Hong-ou-mandel interferometry on a biphoton beat note},\ }\href@noop
  {} {\bibfield  {journal} {\bibinfo  {journal} {npj Quantum Information}\
  }\textbf {\bibinfo {volume} {5}},\ \bibinfo {pages} {43} (\bibinfo {year}
  {2019}{\natexlab{a}})}\BibitemShut {NoStop}%
\bibitem [{\citenamefont {Lipka}\ and\ \citenamefont
  {Parniak}(2021)}]{lipka2021}%
  \BibitemOpen
  \bibfield  {author} {\bibinfo {author} {\bibfnamefont {M.}~\bibnamefont
  {Lipka}}\ and\ \bibinfo {author} {\bibfnamefont {M.}~\bibnamefont
  {Parniak}},\ }\bibfield  {title} {\bibinfo {title} {Single-photon hologram of
  a zero-area pulse},\ }\href@noop {} {\bibfield  {journal} {\bibinfo
  {journal} {Physical Review Letters}\ }\textbf {\bibinfo {volume} {127}},\
  \bibinfo {pages} {163601} (\bibinfo {year} {2021})}\BibitemShut {NoStop}%
\bibitem [{\citenamefont {Fabre}\ and\ \citenamefont
  {Felicetti}(2021)}]{frequencyshift}%
  \BibitemOpen
  \bibfield  {author} {\bibinfo {author} {\bibfnamefont {N.}~\bibnamefont
  {Fabre}}\ and\ \bibinfo {author} {\bibfnamefont {S.}~\bibnamefont
  {Felicetti}},\ }\bibfield  {title} {\bibinfo {title} {Parameter estimation of
  time and frequency shifts with generalized hong-ou-mandel interferometry},\
  }\href {https://doi.org/10.1103/PhysRevA.104.022208} {\bibfield  {journal}
  {\bibinfo  {journal} {Phys. Rev. A}\ }\textbf {\bibinfo {volume} {104}},\
  \bibinfo {pages} {022208} (\bibinfo {year} {2021})}\BibitemShut {NoStop}%
\bibitem [{\citenamefont {Restuccia}\ \emph {et~al.}(2019)\citenamefont
  {Restuccia}, \citenamefont {Toro{\v{s}}}, \citenamefont {Gibson},
  \citenamefont {Ulbricht}, \citenamefont {Faccio},\ and\ \citenamefont
  {Padgett}}]{Photongyroscope}%
  \BibitemOpen
  \bibfield  {author} {\bibinfo {author} {\bibfnamefont {S.}~\bibnamefont
  {Restuccia}}, \bibinfo {author} {\bibfnamefont {M.}~\bibnamefont
  {Toro{\v{s}}}}, \bibinfo {author} {\bibfnamefont {G.~M.}\ \bibnamefont
  {Gibson}}, \bibinfo {author} {\bibfnamefont {H.}~\bibnamefont {Ulbricht}},
  \bibinfo {author} {\bibfnamefont {D.}~\bibnamefont {Faccio}},\ and\ \bibinfo
  {author} {\bibfnamefont {M.~J.}\ \bibnamefont {Padgett}},\ }\bibfield
  {title} {\bibinfo {title} {Photon bunching in a rotating reference frame},\
  }\href@noop {} {\bibfield  {journal} {\bibinfo  {journal} {Physical Review
  Letters}\ }\textbf {\bibinfo {volume} {123}},\ \bibinfo {pages} {110401}
  (\bibinfo {year} {2019})}\BibitemShut {NoStop}%
\bibitem [{\citenamefont {Ndagano}\ \emph {et~al.}(2022)\citenamefont
  {Ndagano}, \citenamefont {Defienne}, \citenamefont {Branford}, \citenamefont
  {Shah}, \citenamefont {Lyons}, \citenamefont {Westerberg}, \citenamefont
  {Gauger},\ and\ \citenamefont {Faccio}}]{quantummicroscopy}%
  \BibitemOpen
  \bibfield  {author} {\bibinfo {author} {\bibfnamefont {B.}~\bibnamefont
  {Ndagano}}, \bibinfo {author} {\bibfnamefont {H.}~\bibnamefont {Defienne}},
  \bibinfo {author} {\bibfnamefont {D.}~\bibnamefont {Branford}}, \bibinfo
  {author} {\bibfnamefont {Y.~D.}\ \bibnamefont {Shah}}, \bibinfo {author}
  {\bibfnamefont {A.}~\bibnamefont {Lyons}}, \bibinfo {author} {\bibfnamefont
  {N.}~\bibnamefont {Westerberg}}, \bibinfo {author} {\bibfnamefont {E.~M.}\
  \bibnamefont {Gauger}},\ and\ \bibinfo {author} {\bibfnamefont
  {D.}~\bibnamefont {Faccio}},\ }\bibfield  {title} {\bibinfo {title} {Quantum
  microscopy based on hong--ou--mandel interference},\ }\href@noop {}
  {\bibfield  {journal} {\bibinfo  {journal} {Nature Photonics}\ }\textbf
  {\bibinfo {volume} {16}},\ \bibinfo {pages} {384} (\bibinfo {year}
  {2022})}\BibitemShut {NoStop}%
\bibitem [{\citenamefont {Rao}(1945)}]{Rao1945}%
  \BibitemOpen
  \bibfield  {author} {\bibinfo {author} {\bibfnamefont {C.~R.}\ \bibnamefont
  {Rao}},\ }\bibfield  {title} {\bibinfo {title} {Information and the accuracy
  attainable in the estimation of statistical parameters},\ }\href@noop {}
  {\bibfield  {journal} {\bibinfo  {journal} {Reson. J. Sci. Educ}\ }\textbf
  {\bibinfo {volume} {20}},\ \bibinfo {pages} {78} (\bibinfo {year}
  {1945})}\BibitemShut {NoStop}%
\bibitem [{\citenamefont {Cram{\'e}r}(1999)}]{cramer1946}%
  \BibitemOpen
  \bibfield  {author} {\bibinfo {author} {\bibfnamefont {H.}~\bibnamefont
  {Cram{\'e}r}},\ }\href@noop {} {\emph {\bibinfo {title} {Mathematical methods
  of statistics}}},\ Vol.~\bibinfo {volume} {26}\ (\bibinfo  {publisher}
  {Princeton university press},\ \bibinfo {year} {1999})\BibitemShut {NoStop}%
\bibitem [{\citenamefont {Harnchaiwat}\ \emph {et~al.}(2020)\citenamefont
  {Harnchaiwat}, \citenamefont {Zhu}, \citenamefont {Westerberg}, \citenamefont
  {Gauger},\ and\ \citenamefont {Leach}}]{harnchaiwat2020tracking}%
  \BibitemOpen
  \bibfield  {author} {\bibinfo {author} {\bibfnamefont {N.}~\bibnamefont
  {Harnchaiwat}}, \bibinfo {author} {\bibfnamefont {F.}~\bibnamefont {Zhu}},
  \bibinfo {author} {\bibfnamefont {N.}~\bibnamefont {Westerberg}}, \bibinfo
  {author} {\bibfnamefont {E.}~\bibnamefont {Gauger}},\ and\ \bibinfo {author}
  {\bibfnamefont {J.}~\bibnamefont {Leach}},\ }\bibfield  {title} {\bibinfo
  {title} {Tracking the polarisation state of light via hong-ou-mandel
  interferometry},\ }\href@noop {} {\bibfield  {journal} {\bibinfo  {journal}
  {Optics express}\ }\textbf {\bibinfo {volume} {28}},\ \bibinfo {pages} {2210}
  (\bibinfo {year} {2020})}\BibitemShut {NoStop}%
\bibitem [{\citenamefont {Johnson}\ \emph {et~al.}(2023)\citenamefont
  {Johnson}, \citenamefont {Lualdi}, \citenamefont {Conrad}, \citenamefont
  {Arnold}, \citenamefont {Vayninger},\ and\ \citenamefont
  {Kwiat}}]{kwait2023}%
  \BibitemOpen
  \bibfield  {author} {\bibinfo {author} {\bibfnamefont {S.~J.}\ \bibnamefont
  {Johnson}}, \bibinfo {author} {\bibfnamefont {C.~P.}\ \bibnamefont {Lualdi}},
  \bibinfo {author} {\bibfnamefont {A.~P.}\ \bibnamefont {Conrad}}, \bibinfo
  {author} {\bibfnamefont {N.~T.}\ \bibnamefont {Arnold}}, \bibinfo {author}
  {\bibfnamefont {M.}~\bibnamefont {Vayninger}},\ and\ \bibinfo {author}
  {\bibfnamefont {P.~G.}\ \bibnamefont {Kwiat}},\ }\bibfield  {title} {\bibinfo
  {title} {Toward vibration measurement via frequency-entangled two-photon
  interferometry},\ }in\ \href@noop {} {\emph {\bibinfo {booktitle} {Quantum
  Sensing, Imaging, and Precision Metrology}}},\ Vol.\ \bibinfo {volume}
  {12447}\ (\bibinfo {organization} {SPIE},\ \bibinfo {year} {2023})\ pp.\
  \bibinfo {pages} {263--268}\BibitemShut {NoStop}%
\bibitem [{\citenamefont {Jordan}\ \emph {et~al.}(2022)\citenamefont {Jordan},
  \citenamefont {Abrahao},\ and\ \citenamefont {Lundeen}}]{timinglimits}%
  \BibitemOpen
  \bibfield  {author} {\bibinfo {author} {\bibfnamefont {K.~M.}\ \bibnamefont
  {Jordan}}, \bibinfo {author} {\bibfnamefont {R.~A.}\ \bibnamefont
  {Abrahao}},\ and\ \bibinfo {author} {\bibfnamefont {J.~S.}\ \bibnamefont
  {Lundeen}},\ }\bibfield  {title} {\bibinfo {title} {Quantum metrology timing
  limits of the hong-ou-mandel interferometer and of general two-photon
  measurements},\ }\href@noop {} {\bibfield  {journal} {\bibinfo  {journal}
  {Physical Review A}\ }\textbf {\bibinfo {volume} {106}},\ \bibinfo {pages}
  {063715} (\bibinfo {year} {2022})}\BibitemShut {NoStop}%
\bibitem [{\citenamefont {Triggiani}\ \emph {et~al.}(2023)\citenamefont
  {Triggiani}, \citenamefont {Psaroudis},\ and\ \citenamefont
  {Tamma}}]{tamma2023}%
  \BibitemOpen
  \bibfield  {author} {\bibinfo {author} {\bibfnamefont {D.}~\bibnamefont
  {Triggiani}}, \bibinfo {author} {\bibfnamefont {G.}~\bibnamefont
  {Psaroudis}},\ and\ \bibinfo {author} {\bibfnamefont {V.}~\bibnamefont
  {Tamma}},\ }\bibfield  {title} {\bibinfo {title} {Ultimate quantum
  sensitivity in the estimation of the delay between two interfering photons
  through frequency-resolving sampling},\ }\href@noop {} {\bibfield  {journal}
  {\bibinfo  {journal} {Physical Review Applied}\ }\textbf {\bibinfo {volume}
  {19}},\ \bibinfo {pages} {044068} (\bibinfo {year} {2023})}\BibitemShut
  {NoStop}%
\bibitem [{\citenamefont {Scott}\ \emph {et~al.}(2020)\citenamefont {Scott},
  \citenamefont {Branford}, \citenamefont {Westerberg}, \citenamefont {Leach},\
  and\ \citenamefont {Gauger}}]{Beyondcoincidence}%
  \BibitemOpen
  \bibfield  {author} {\bibinfo {author} {\bibfnamefont {H.}~\bibnamefont
  {Scott}}, \bibinfo {author} {\bibfnamefont {D.}~\bibnamefont {Branford}},
  \bibinfo {author} {\bibfnamefont {N.}~\bibnamefont {Westerberg}}, \bibinfo
  {author} {\bibfnamefont {J.}~\bibnamefont {Leach}},\ and\ \bibinfo {author}
  {\bibfnamefont {E.~M.}\ \bibnamefont {Gauger}},\ }\bibfield  {title}
  {\bibinfo {title} {Beyond coincidence in hong-ou-mandel interferometry},\
  }\href {https://doi.org/10.1103/PhysRevA.102.033714} {\bibfield  {journal}
  {\bibinfo  {journal} {Phys. Rev. A}\ }\textbf {\bibinfo {volume} {102}},\
  \bibinfo {pages} {033714} (\bibinfo {year} {2020})}\BibitemShut {NoStop}%
\bibitem [{\citenamefont {Scott}\ \emph {et~al.}(2021)\citenamefont {Scott},
  \citenamefont {Branford}, \citenamefont {Westerberg}, \citenamefont {Leach},\
  and\ \citenamefont {Gauger}}]{Noiselimit}%
  \BibitemOpen
  \bibfield  {author} {\bibinfo {author} {\bibfnamefont {H.}~\bibnamefont
  {Scott}}, \bibinfo {author} {\bibfnamefont {D.}~\bibnamefont {Branford}},
  \bibinfo {author} {\bibfnamefont {N.}~\bibnamefont {Westerberg}}, \bibinfo
  {author} {\bibfnamefont {J.}~\bibnamefont {Leach}},\ and\ \bibinfo {author}
  {\bibfnamefont {E.~M.}\ \bibnamefont {Gauger}},\ }\bibfield  {title}
  {\bibinfo {title} {Noise limits on two-photon interferometric sensing},\
  }\href {https://doi.org/10.1103/PhysRevA.104.053704} {\bibfield  {journal}
  {\bibinfo  {journal} {Phys. Rev. A}\ }\textbf {\bibinfo {volume} {104}},\
  \bibinfo {pages} {053704} (\bibinfo {year} {2021})}\BibitemShut {NoStop}%
\bibitem [{\citenamefont {Chen}\ \emph
  {et~al.}(2019{\natexlab{b}})\citenamefont {Chen}, \citenamefont {Fink},
  \citenamefont {Steinlechner}, \citenamefont {Torres},\ and\ \citenamefont
  {Ursin}}]{beatnote_biphoton}%
  \BibitemOpen
  \bibfield  {author} {\bibinfo {author} {\bibfnamefont {Y.}~\bibnamefont
  {Chen}}, \bibinfo {author} {\bibfnamefont {M.}~\bibnamefont {Fink}}, \bibinfo
  {author} {\bibfnamefont {F.}~\bibnamefont {Steinlechner}}, \bibinfo {author}
  {\bibfnamefont {J.~P.}\ \bibnamefont {Torres}},\ and\ \bibinfo {author}
  {\bibfnamefont {R.}~\bibnamefont {Ursin}},\ }\bibfield  {title} {\bibinfo
  {title} {Hong-ou-mandel interferometry on a biphoton beat note},\ }\href@noop
  {} {\bibfield  {journal} {\bibinfo  {journal} {npj Quantum Information}\
  }\textbf {\bibinfo {volume} {5}},\ \bibinfo {pages} {43} (\bibinfo {year}
  {2019}{\natexlab{b}})}\BibitemShut {NoStop}%
\bibitem [{\citenamefont {Klyshko}(1967)}]{klyshko1967coherent}%
  \BibitemOpen
  \bibfield  {author} {\bibinfo {author} {\bibfnamefont {D.}~\bibnamefont
  {Klyshko}},\ }\bibfield  {title} {\bibinfo {title} {Coherent photon decay in
  a nonlinear medium},\ }\href@noop {} {\bibfield  {journal} {\bibinfo
  {journal} {ZhETF Pisma Redaktsiiu}\ }\textbf {\bibinfo {volume} {6}},\
  \bibinfo {pages} {490} (\bibinfo {year} {1967})}\BibitemShut {NoStop}%
\bibitem [{\citenamefont {Boyd}(2020)}]{boyd2020nonlinear}%
  \BibitemOpen
  \bibfield  {author} {\bibinfo {author} {\bibfnamefont {R.~W.}\ \bibnamefont
  {Boyd}},\ }\href@noop {} {\emph {\bibinfo {title} {Nonlinear optics}}}\
  (\bibinfo  {publisher} {Academic press},\ \bibinfo {year} {2020})\BibitemShut
  {NoStop}%
\bibitem [{\citenamefont {Ebrahimzadeh}\ and\ \citenamefont
  {Dunn}(2001)}]{ebrahimzadeh2001optical}%
  \BibitemOpen
  \bibfield  {author} {\bibinfo {author} {\bibfnamefont {M.}~\bibnamefont
  {Ebrahimzadeh}}\ and\ \bibinfo {author} {\bibfnamefont {M.}~\bibnamefont
  {Dunn}},\ }\bibfield  {title} {\bibinfo {title} {Optical parametric
  oscillators},\ }\href@noop {} {\bibfield  {journal} {\bibinfo  {journal} {OSA
  Handbook of Optics}\ }\textbf {\bibinfo {volume} {4}},\ \bibinfo {pages} {22}
  (\bibinfo {year} {2001})}\BibitemShut {NoStop}%
\bibitem [{\citenamefont {Okano}\ \emph {et~al.}(2015)\citenamefont {Okano},
  \citenamefont {Lim}, \citenamefont {Okamoto}, \citenamefont {Nishizawa},
  \citenamefont {Kurimura},\ and\ \citenamefont
  {Takeuchi}}]{okano201500.5micron}%
  \BibitemOpen
  \bibfield  {author} {\bibinfo {author} {\bibfnamefont {M.}~\bibnamefont
  {Okano}}, \bibinfo {author} {\bibfnamefont {H.~H.}\ \bibnamefont {Lim}},
  \bibinfo {author} {\bibfnamefont {R.}~\bibnamefont {Okamoto}}, \bibinfo
  {author} {\bibfnamefont {N.}~\bibnamefont {Nishizawa}}, \bibinfo {author}
  {\bibfnamefont {S.}~\bibnamefont {Kurimura}},\ and\ \bibinfo {author}
  {\bibfnamefont {S.}~\bibnamefont {Takeuchi}},\ }\bibfield  {title} {\bibinfo
  {title} {0.54 $\mu$m resolution two-photon interference with dispersion
  cancellation for quantum optical coherence tomography},\ }\href@noop {}
  {\bibfield  {journal} {\bibinfo  {journal} {Scientific reports}\ }\textbf
  {\bibinfo {volume} {5}},\ \bibinfo {pages} {18042} (\bibinfo {year}
  {2015})}\BibitemShut {NoStop}%
\bibitem [{\citenamefont {Jabir}\ and\ \citenamefont
  {Samanta}(2017)}]{Jabir:17}%
  \BibitemOpen
  \bibfield  {author} {\bibinfo {author} {\bibfnamefont {M.~V.}\ \bibnamefont
  {Jabir}}\ and\ \bibinfo {author} {\bibfnamefont {G.~K.}\ \bibnamefont
  {Samanta}},\ }\bibfield  {title} {\bibinfo {title} {{Robust, high brightness,
  degenerate entangled photon source at room temperature}},\ }\href
  {https://doi.org/10.1038/s41598-017-12709-5} {\bibfield  {journal} {\bibinfo
  {journal} {Scientific Reports}\ }\textbf {\bibinfo {volume} {7}},\ \bibinfo
  {pages} {12613} (\bibinfo {year} {2017})}\BibitemShut {NoStop}%
\bibitem [{\citenamefont {Singh}\ \emph {et~al.}(2022)\citenamefont {Singh},
  \citenamefont {Kumar}, \citenamefont {Ghosh}, \citenamefont {Forbes},\ and\
  \citenamefont {Samanta}}]{singh2023}%
  \BibitemOpen
  \bibfield  {author} {\bibinfo {author} {\bibfnamefont {S.}~\bibnamefont
  {Singh}}, \bibinfo {author} {\bibfnamefont {V.}~\bibnamefont {Kumar}},
  \bibinfo {author} {\bibfnamefont {A.}~\bibnamefont {Ghosh}}, \bibinfo
  {author} {\bibfnamefont {A.}~\bibnamefont {Forbes}},\ and\ \bibinfo {author}
  {\bibfnamefont {G.~K.}\ \bibnamefont {Samanta}},\ }\bibfield  {title}
  {\bibinfo {title} {A tolerance-enhanced spontaneous parametric downconversion
  source of bright entangled photons},\ }\href
  {https://doi.org/https://doi.org/10.1002/qute.202200121} {\bibfield
  {journal} {\bibinfo  {journal} {Advanced Quantum Technologies}\ }\textbf
  {\bibinfo {volume} {6}},\ \bibinfo {pages} {2200121} (\bibinfo {year}
  {2022})}\BibitemShut {NoStop}%
\bibitem [{\citenamefont {Das}\ \emph {et~al.}(2009)\citenamefont {Das},
  \citenamefont {Kumar}, \citenamefont {Samanta},\ and\ \citenamefont
  {Ebrahim-Zadeh}}]{RDas:09}%
  \BibitemOpen
  \bibfield  {author} {\bibinfo {author} {\bibfnamefont {R.}~\bibnamefont
  {Das}}, \bibinfo {author} {\bibfnamefont {S.~C.}\ \bibnamefont {Kumar}},
  \bibinfo {author} {\bibfnamefont {G.~K.}\ \bibnamefont {Samanta}},\ and\
  \bibinfo {author} {\bibfnamefont {M.}~\bibnamefont {Ebrahim-Zadeh}},\
  }\bibfield  {title} {\bibinfo {title} {Broadband, high-power,
  continuous-wave, mid-infrared source using extended phase-matching bandwidth
  in mgo:ppln},\ }\href {https://doi.org/10.1364/OL.34.003836} {\bibfield
  {journal} {\bibinfo  {journal} {Opt. Lett.}\ }\textbf {\bibinfo {volume}
  {34}},\ \bibinfo {pages} {3836} (\bibinfo {year} {2009})}\BibitemShut
  {NoStop}%
\bibitem [{\citenamefont {Chen}\ \emph {et~al.}(2021)\citenamefont {Chen},
  \citenamefont {Xu}, \citenamefont {Riazi}, \citenamefont {Zhu}, \citenamefont
  {Gladyshev}, \citenamefont {Kazansky},\ and\ \citenamefont
  {Qian}}]{chen2021broadband}%
  \BibitemOpen
  \bibfield  {author} {\bibinfo {author} {\bibfnamefont {C.}~\bibnamefont
  {Chen}}, \bibinfo {author} {\bibfnamefont {C.}~\bibnamefont {Xu}}, \bibinfo
  {author} {\bibfnamefont {A.}~\bibnamefont {Riazi}}, \bibinfo {author}
  {\bibfnamefont {E.~Y.}\ \bibnamefont {Zhu}}, \bibinfo {author} {\bibfnamefont
  {A.~V.}\ \bibnamefont {Gladyshev}}, \bibinfo {author} {\bibfnamefont {P.~G.}\
  \bibnamefont {Kazansky}},\ and\ \bibinfo {author} {\bibfnamefont
  {L.}~\bibnamefont {Qian}},\ }\bibfield  {title} {\bibinfo {title} {Broadband
  fiber-based entangled photon-pair source at telecom o-band},\ }\href@noop {}
  {\bibfield  {journal} {\bibinfo  {journal} {Optics Letters}\ }\textbf
  {\bibinfo {volume} {46}},\ \bibinfo {pages} {1261} (\bibinfo {year}
  {2021})}\BibitemShut {NoStop}%
\bibitem [{\citenamefont {Abouraddy}\ \emph {et~al.}(2002)\citenamefont
  {Abouraddy}, \citenamefont {Nasr}, \citenamefont {Saleh}, \citenamefont
  {Sergienko},\ and\ \citenamefont {Teich}}]{abouraddy2002QOCT}%
  \BibitemOpen
  \bibfield  {author} {\bibinfo {author} {\bibfnamefont {A.~F.}\ \bibnamefont
  {Abouraddy}}, \bibinfo {author} {\bibfnamefont {M.~B.}\ \bibnamefont {Nasr}},
  \bibinfo {author} {\bibfnamefont {B.~E.}\ \bibnamefont {Saleh}}, \bibinfo
  {author} {\bibfnamefont {A.~V.}\ \bibnamefont {Sergienko}},\ and\ \bibinfo
  {author} {\bibfnamefont {M.~C.}\ \bibnamefont {Teich}},\ }\bibfield  {title}
  {\bibinfo {title} {Quantum-optical coherence tomography with dispersion
  cancellation},\ }\href@noop {} {\bibfield  {journal} {\bibinfo  {journal}
  {Physical Review A}\ }\textbf {\bibinfo {volume} {65}},\ \bibinfo {pages}
  {053817} (\bibinfo {year} {2002})}\BibitemShut {NoStop}%
\bibitem [{\citenamefont {Ham}(2022)}]{BSham2022coherence}%
  \BibitemOpen
  \bibfield  {author} {\bibinfo {author} {\bibfnamefont {B.~S.}\ \bibnamefont
  {Ham}},\ }\bibfield  {title} {\bibinfo {title} {Coherence interpretation of
  the hong-ou-mandel effect},\ }\href@noop {} {\bibfield  {journal} {\bibinfo
  {journal} {arXiv preprint arXiv:2203.13983}\ } (\bibinfo {year}
  {2022})}\BibitemShut {NoStop}%
\bibitem [{\citenamefont {Yepiz-Graciano}\ \emph {et~al.}(2020)\citenamefont
  {Yepiz-Graciano}, \citenamefont {Mart{\'\i}nez}, \citenamefont {Lopez-Mago},
  \citenamefont {Cruz-Ramirez},\ and\ \citenamefont
  {U’Ren}}]{2020spectrallyresolved}%
  \BibitemOpen
  \bibfield  {author} {\bibinfo {author} {\bibfnamefont {P.}~\bibnamefont
  {Yepiz-Graciano}}, \bibinfo {author} {\bibfnamefont {A.~M.~A.}\ \bibnamefont
  {Mart{\'\i}nez}}, \bibinfo {author} {\bibfnamefont {D.}~\bibnamefont
  {Lopez-Mago}}, \bibinfo {author} {\bibfnamefont {H.}~\bibnamefont
  {Cruz-Ramirez}},\ and\ \bibinfo {author} {\bibfnamefont {A.~B.}\ \bibnamefont
  {U’Ren}},\ }\bibfield  {title} {\bibinfo {title} {Spectrally resolved
  hong--ou--mandel interferometry for quantum-optical coherence tomography},\
  }\href@noop {} {\bibfield  {journal} {\bibinfo  {journal} {Photonics
  Research}\ }\textbf {\bibinfo {volume} {8}},\ \bibinfo {pages} {1023}
  (\bibinfo {year} {2020})}\BibitemShut {NoStop}%
\bibitem [{\citenamefont {Nasr}\ \emph {et~al.}(2003)\citenamefont {Nasr},
  \citenamefont {Saleh}, \citenamefont {Sergienko},\ and\ \citenamefont
  {Teich}}]{PhysRevLett.91.083601}%
  \BibitemOpen
  \bibfield  {author} {\bibinfo {author} {\bibfnamefont {M.~B.}\ \bibnamefont
  {Nasr}}, \bibinfo {author} {\bibfnamefont {B.~E.~A.}\ \bibnamefont {Saleh}},
  \bibinfo {author} {\bibfnamefont {A.~V.}\ \bibnamefont {Sergienko}},\ and\
  \bibinfo {author} {\bibfnamefont {M.~C.}\ \bibnamefont {Teich}},\ }\bibfield
  {title} {\bibinfo {title} {Demonstration of dispersion-canceled
  quantum-optical coherence tomography},\ }\href
  {https://doi.org/10.1103/PhysRevLett.91.083601} {\bibfield  {journal}
  {\bibinfo  {journal} {Phys. Rev. Lett.}\ }\textbf {\bibinfo {volume} {91}},\
  \bibinfo {pages} {083601} (\bibinfo {year} {2003})}\BibitemShut {NoStop}%
\bibitem [{\citenamefont {Zhou}\ \emph {et~al.}(2012)\citenamefont {Zhou},
  \citenamefont {Gao}, \citenamefont {Yang},\ and\ \citenamefont
  {Tian}}]{zhou2012piezo}%
  \BibitemOpen
  \bibfield  {author} {\bibinfo {author} {\bibfnamefont {M.}~\bibnamefont
  {Zhou}}, \bibinfo {author} {\bibfnamefont {W.}~\bibnamefont {Gao}}, \bibinfo
  {author} {\bibfnamefont {Z.}~\bibnamefont {Yang}},\ and\ \bibinfo {author}
  {\bibfnamefont {Y.}~\bibnamefont {Tian}},\ }\bibfield  {title} {\bibinfo
  {title} {High precise fuzzy control for piezoelectric direct drive
  electro-hydraulic servo valve},\ }\href@noop {} {\bibfield  {journal}
  {\bibinfo  {journal} {Journal of Advanced Mechanical Design, Systems, and
  Manufacturing}\ }\textbf {\bibinfo {volume} {6}},\ \bibinfo {pages} {1154}
  (\bibinfo {year} {2012})}\BibitemShut {NoStop}%
\bibitem [{\citenamefont {Grzybek}\ and\ \citenamefont
  {Sioma}(2022)}]{grzybek2022creep}%
  \BibitemOpen
  \bibfield  {author} {\bibinfo {author} {\bibfnamefont {D.}~\bibnamefont
  {Grzybek}}\ and\ \bibinfo {author} {\bibfnamefont {A.}~\bibnamefont
  {Sioma}},\ }\bibfield  {title} {\bibinfo {title} {Creep phenomenon in a
  multiple-input single-output control system of a piezoelectric bimorph
  actuator},\ }\href@noop {} {\bibfield  {journal} {\bibinfo  {journal}
  {Energies}\ }\textbf {\bibinfo {volume} {15}},\ \bibinfo {pages} {8267}
  (\bibinfo {year} {2022})}\BibitemShut {NoStop}%
\bibitem [{\citenamefont {Fox}\ and\ \citenamefont
  {Fox}(2006)}]{fox2006quantum}%
  \BibitemOpen
  \bibfield  {author} {\bibinfo {author} {\bibfnamefont {A.~M.}\ \bibnamefont
  {Fox}}\ and\ \bibinfo {author} {\bibfnamefont {M.}~\bibnamefont {Fox}},\
  }\href@noop {} {\emph {\bibinfo {title} {Quantum optics: an introduction}}},\
  Vol.~\bibinfo {volume} {15}\ (\bibinfo  {publisher} {Oxford university
  press},\ \bibinfo {year} {2006})\BibitemShut {NoStop}%
\end{thebibliography}%

\end{document}